\documentstyle[preprint,prb,aps,epsf]{revtex}
\begin{document}

\draft

\title{
Extended Dynamical Mean Field Theory
and GW method
}
\author{Ping Sun and Gabriel Kotliar}
\address{
Center for Materials Theory,
Department of Physics and Astronomy,
Rutgers University,
Piscataway, NJ 08854-8019
}

\date{\today}

\maketitle

\begin{abstract}
We develop the extended dynamical mean field theory (E-DMFT) with
a view towards realistic applications.
{\bf 1)} We introduce an intuitive derivation of the E-DMFT formalism.
By identifying the Hartree contributions before the E-DMFT treatment,
it allows to handle systems in symmetry breaking phases within a simple
formalism.
{\bf 2)} We make a new implementation of E-DMFT through real
Hubbard-Stratonovich
transformation to decouple the non-local two-particle interactions. 
We apply it to a 3D U-V model, with U the on-site and V the nearest neighbor
interactions,  and investigate the behavior of the
various Green's functions, especially the density susceptibility,
as the density instability is approached. We obtain the phase
diagram at a finite temperature.
{\bf 3)} We present a formalism incorporating E-DMFT with Cellular DMFT.
{\bf 4)} We suggest an improvement of the E-DMFT approach by combining
it with a generalized GW method. The method combines the local
self-energy from E-DMFT and the non-local ones from the perturbative
calculation of GW. We apply the method to a 1D U-V model with two
sublattices carrying different chemical potentials. By comparing with
those from Density Matrix Renormalization Group (DMRG) calculations, we
show the results are shifted in the correct direction due to the GW
contributions. {\bf 5)} In order to handle the
generic Coulomb repulsion within E-DMFT, we describe a method to
tailor E-DMFT so that proper momentum dependence can be kept in
general response functions.

\end{abstract}


\section{Introduction}
\label{sec-01}

The Dynamical Mean Field Theory (DMFT) has been a powerful tool
for the study of strongly correlated electronic systems. It has
allowed us to gain new insights into non-perturbative problems
such as the Mott paramagnetic metal to paramagnetic insulator
transition at finite temperatures \cite{kotliar1996}. There are
many current attempts to extend the scope of the DMFT approach
in several directions: to include momentum dependence in the
self-energy \cite{kotliar1996,dongen,schiller,jarrell,georges,kotliar2001,kotliar2002},
to account for the effects of finite range interactions
\cite{subir,kajueter,si2000,georges1999,chitra2000,motome,pankov},
and to combine DMFT with realistic band structure
\cite{anisimov,savrasov2001,gabi,held}.\\

The current paper is devoted to the investigation of the Extended
Dynamical Mean Field Theory (E-DMFT), an extension of the original
DMFT in order to take into account the spatially non-local
interactions beyond the Hartree level. The idea of E-DMFT was
developed independently in the studies of spin glass
\cite{subir,georges1999}, systems with non-local Coulomb interaction
\cite{kajueter,chitra2000}, and the heavy fermion system \cite{si2000}.
The derivation of E-DMFT based on Baym-Kadanoff functional has been
achieved in Ref.[\cite{chitra2000}].\\

We present here several methodological developments which build on the
E-DMFT approach with a view to
obtain a more realistic description of solids. Our goals are to
describe {\bf a)} the frequency dependence of the effective
interaction and its effects on the single particle spectra,
{\bf b)} the effects of short range correlations, and {\bf c)}
a momentum dependent polarization. These three
effects certainly present in realistic models of solids
\cite{hirsch2001,ferdi,zein,mahan}. We discuss them in this
paper in the framework of model Hamiltonians in sections \ref{sec-04},
\ref{sec-07}, and \ref{sec-08}. It should be pointed out that the problem
is not touched in this paper as how to obtain the parameters of the
model Hamiltonian from the first principle calculations.\\

In addition we present several technical advances for the analysis
of E-DMFT equations. In sections \ref{sec-02} and \ref{sec-03}, we
present a
simplified derivation of these equations in a broken symmetry phase.
A method for handling arbitrary interactions within E-DMFT-QMC by
an interaction shift \cite{hamann} is also discussed in the section.
We show the formalism combining E-DMFT with cellular DMFT (C-DMFT) in
\S\ref{sec-05} and with GW approximation \cite{hedin} in
\S\ref{sec-06}.\\

To illustrate the ideas and the working of the methods we apply
them to two problems. The transitions between a Mott insulator
(MI), a band insulator (BI), and a Fermi liquid (FL) in a 3D U-V
model is discussed in \S\ref{sec-04}. The model describes an
electron system with an on-site repulsion U and a nearest neighbor
interaction V. It was treated in
simple DMFT at half \cite{dongen} and quarter \cite{pietig}
fillings and is relevant to the materials with charge ordered
phase \cite{c-order1,c-order2,c-order3}. The problem of the
transition between MI and BI phases in a 1D U-V model with
alternating chemical potential was
discussed in the context of mixed stack organic compounds
\cite{torrance,nagaosa} and ferroelectric perovskites \cite{egami}.
We exhibit in \S\ref{sec-07} the implementation of E-DMFT-GW method
on the 1D model. \S\ref{sec-09} is the conclusion.\\

\section{Model Hamiltonian}
\label{sec-02}

We start with the following Hamiltonian:

\[
  \hat{H}= -\frac{1}{2}\sum_{i,j}\sum_{\sigma} (t_{ij}
  \hat{c}_{i\sigma}^{\dagger}\hat{c}_{j\sigma}+\; {\rm h.c.})
  - \sum_{i}\sum_{\alpha=0}^3 h_{i\alpha}\hat{S}_{i\alpha}
\]
\begin{equation}
\label{eq-01}
  +U\sum_i \hat{n}_{i\uparrow}\hat{n}_{i\downarrow}
  +\frac{1}{2}\sum_{i,j}\sum_{\alpha,\beta=0}^{3} 
  \hat{S}_{i\alpha} V_{i\alpha,j\beta}\hat{S}_{j\beta}.
\end{equation}

\noindent The operators $\hat{S}_{i\alpha}\stackrel{\rm def}{=}
\hat{c}_{i\sigma}^{\dagger} \tau^{\alpha}_{\sigma,\sigma^{\prime}}
\hat{c}_{i\sigma^{\prime}}$ with $\tau^{\alpha}$ being the Pauli
matrices for $\alpha=1,2,3$\ and $\tau^{0}=I_{2 \times 2}$, the
identity. So the zero-th component of $\hat{S}_{i\alpha}$ is the
charge density and the rest the spin operators.
Similarly, $h_{i\alpha=0}$ represents the chemical potential
and the other three, with $\alpha=1,2,3$, the external magnetic
field.\\

In Eq.(\ref{eq-01}) the on-site part of the interaction is isolated
in $U$ and the off-site parts are described by $V_{i\alpha,j\beta}$,
hence $V_{i\alpha,i\beta}=0$. We have allowed all the possible forms
of the instantaneous direct and exchange interactions, but excluded
some others, like the pair hoppings. Unless otherwise specified, the
form of the interaction $V_{i\alpha,j\beta}$ will be generic with
only the simple requirements that the interaction be translational
invariant, $V_{i\alpha,j\beta}=V_{\alpha,\beta}(|i-j|)$, and
symmetric, $V_{i\alpha,j\beta}=V_{j\beta,i\alpha}$. Due to the
possible ionic screening and the super-exchange mechanisms, we have
the freedom to certain extent in choosing $U$ and the components of
$V$ independently at the level of the model Hamiltonian. While it is
reasonable to take $U>0$, in the off-site interaction matrix $V$ all
possibilities are allowed if some effective forms of the interactions
are under consideration. In other words, the matrix $V$ is not
necessarily positive- or negative- definite.\\

In this paper we keep for simplicity $h_0, h_3 \neq 0$ and set
$h_1, h_2 =0$. This choice is more general than it appears since we
can always choose the direction of the effective magnetic field to be
the z-direction by appropriately rotating the system. Furthermore, we
assume that the spin symmetry breaking, if it may happen, is also
along the z-direction. This assumption is very natural if
$V_{i\alpha,j\beta}$ is symmetric under the permutation among the
three directions. However, it is actually more restrictive since it
requires that if $V_{\alpha,\beta}(|i-j|)$ produces any easy-axis, the
direction of the easy-axis should be also in the z-direction. We can
then define a spin-dependent chemical potential:

\[
  \mu_{i\sigma} \stackrel{\rm def}{=} h_{i0}+\sigma h_{i3}
\]

Before deriving the E-DMFT formalism, we need first to
separate out the Hartree contributions from the interactions. It is
known from the Baym-Kadanoff functional analysis of E-DMFT
\cite{chitra2000} that in the phases with broken symmetry, Hartree
terms contribute to the E-DMFT by shifting the chemical potential
\cite{note1}. In the current case, due to the spin exchanges, this
shift is spin-dependent. It is also known that in the
circumstance of C-DMFT \cite{venky2}, one needs to handle
separately the Hartree contributions from the non-local interaction
across the cluster boundary. With all these motivations, we rewrite
the Hamiltonian as follows:

\[
  \hat{H}= -\frac{1}{2}\sum_{i,j}\sum_{\sigma} (t_{ij}
  \hat{c}_{i\sigma}^{\dagger}\hat{c}_{j\sigma}+\; {\rm h.c.})
  - \sum_{i,\sigma} \mu^{\rm eff}_{i\sigma} n_{i\sigma}
\]
\begin{equation}
\label{eq-02}
  +U\sum_i :\hat{n}_{i\uparrow}::\hat{n}_{i\downarrow}:
  +\frac{1}{2}\sum_{i\neq j}\sum_{\alpha,\beta=0}^{3} 
  :\hat{S}_{i\alpha}: V_{i\alpha,j\beta}:\hat{S}_{j\beta}:
\end{equation}

\noindent where the normal ordering of the operators is defined by
$:{\cal O}:\stackrel{\rm def}{=}{\cal O}-\langle {\cal O} \rangle$
with the average over the ground state. The effective chemical
potential is defined to be

\begin{equation}
\label{eq-03}
  \mu^{\rm eff}_{i\sigma}=\mu_{i\sigma}-\frac{1}{2}U
  -\sum_{j(\neq i)} V_{i0,j0}
  (\langle n_{j\uparrow} \rangle + \langle n_{j\downarrow} \rangle)
  -\sigma\sum_{j(\neq i)} V_{i3,j3}
  (\langle n_{j\uparrow} \rangle - \langle n_{j\downarrow} \rangle)
\end{equation}

\noindent We work with the functional integral representation of the
partition function at finite temperature $0< T=1/\beta< \infty$ 
\cite{negele}. We want to perform a Hubbard-Stratonovich transform
and decouple the $V$-interactions \cite{jarrell}. Since ultimately we
will map the many-body system to
an impurity model and solve the self-consistent impurity problem via
numerical techniques like QMC, it is desired that the
Hubbard-Stratonovich transform be real \cite{hamann}. To this
end, we need first to add the identity matrices
$\lambda_{\alpha\beta}I_{ij}$ to the off-site interaction to ensure
that, for given $\alpha$ and $\beta$, the matrix

\begin{equation}
\label{eq-04}
  [\tilde{V}_{\alpha\beta}]_{ij}\stackrel{\rm def}{=}
  \lambda_{\alpha\beta} {\rm I}_{ij}-[V_{\alpha\beta}]_{ij}
\end{equation}

\noindent is positive-definite. The minus sign in front of the bare
interaction in the above equation is needed for a real
Hubbard-Stratonovich transform.  Practically we can take any
value of $\lambda_{\alpha\beta}$ as long as it is greater than the
biggest eigen value of the matrix $[V_{\alpha\beta}]$ whose
elements are defined by
$[V_{\alpha\beta}]_{ij}\stackrel{\rm def}{=}V_{i\alpha,j\beta}$.
To keep the symmetry, we require
$\lambda_{\alpha\beta}=\lambda_{\beta\alpha}$.
Using Hubbard-Stratonovich transform, one can write the partition
function in terms of the following functional integral:

\begin{equation}
\label{eq-05}
  Z=\int {\cal D}[c_{i,\sigma}^{\dagger}(\tau),c_{i,\sigma}(\tau);
  \phi_{i,\alpha}(\tau)] \exp(-S)
\end{equation}

\noindent with the Euclidean action:

\[
  S=\int_0^{\beta} d\tau \{ \sum_{i,\sigma} c_{i\sigma}^{\dagger}(\tau)
  \partial_{\tau} c_{i\sigma}(\tau)
  -\sum_{i,j}\sum_{\sigma}[t_{ij}c_{i\sigma}^{\dagger}(\tau)
  c_{j\sigma}(\tau)+{\rm h.c.}] - \sum_{i,\sigma} \mu^{\rm eff}_{i,\sigma}
  n_{i,\sigma}
\]
\begin{equation}
\label{eq-06}
  +U^{\rm eff} \sum_i 
  :n_{i\uparrow}(\tau): :n_{i\downarrow}(\tau):
  +\frac{1}{2}\sum_{i,j}\sum_{\alpha,\beta=0}^3
  \phi_{i\alpha}(\tau) [\tilde{V}_{\alpha\beta}]_{ij}^{-1}
  \phi_{j,\beta}(\tau)
  \pm \sum_{i}\sum_{\alpha=0}^3
  \phi_{i\alpha}(\tau) :S_{i\alpha}(\tau):\},
\end{equation}

\noindent where

\[
  U^{\rm eff}=U+\tilde{V}_{i0,i0}-\tilde{V}_{i1,i1}
  -\tilde{V}_{i2,i2}-\tilde{V}_{i3,i3}
  =U+\lambda_{00}-\lambda_{11}-\lambda_{22}-\lambda_{33}
\]

\noindent While the $\lambda$ dependences
are explicitly contained in the effective on-site interaction
$U^{\rm eff}$ and the interaction matrices $\tilde{V}_{\alpha\beta}$,
they cancel each other exactly so that all the physically measurable
quantities are independent of $\lambda$. The electron-phonon vertex
defined in Eq.(\ref{eq-06}) is local. In an insulator with a
long range Coulomb interaction, one needs a different choice of the
vertex, as will be discussed later in \S\ref{sec-08}.\\

Several remarks are in place. First, since our strategy is to separate
out all the Hartree contributions from the interactions, we certainly
need to treat the auxiliary field in the same way. However, due to the
identity

\[
  \langle \phi_{i\alpha} \rangle
  = \mp \sum_{j}\sum_{\beta=0}^3 [\tilde{V}_{\alpha\beta}]_{ij}
  \langle :S_{j\beta}: \rangle = 0,
\]

\noindent we need not bother to normal order the $\phi$ fields here.
The current formalism of E-DMFT can be extended directly to a general
bose field \cite{motome} which can develop non-zero
expectation values, if we replace $\phi_{i\alpha}$ by
$:\phi_{i\alpha}:$. Second, there is an arbitrariness
in choosing the sign of the electron-phonon vertex which is
reflected in the ``$\pm$'' signs appeared in Eq.(\ref{eq-06}). It
comes from the freedom in the Ising ($Z_2$) symmetry of the order
parameters. For definiteness, we take `$-$' from now on. Finally,
one can also try to decouple the on-site interaction by using yet
another continuous auxiliary field. Because this interaction does not
get any renormalization as we derive the E-DMFT, and because for the
local density-density interaction there is the more efficient Hirsch-Fye
algorithm \cite{hirsch} which uses the discrete Ising auxiliary field,
we will not implement this here \cite{buendia1}. \\

For future reference, we first define the following Matsubara
Green's function:

\begin{equation}
\label{eq-07}
  G_{i\sigma,j\sigma^{\prime}}(\tau-\tau^{\prime})
  \stackrel{\rm def}{=}
  -\langle {\rm T}_{\tau} c_{i\sigma}(\tau)
  c^{\dagger}_{j\sigma^{\prime}}(\tau^{\prime})\rangle,
\end{equation}
\begin{equation}
\label{eq-08}
  \chi_{i\alpha,j\beta}(\tau-\tau^{\prime})
  \stackrel{\rm def}{=} -\langle {\rm T}_{\tau}
  :S_{i\alpha}(\tau): :S_{j\beta}(\tau^{\prime}):\rangle,
\end{equation}
\begin{equation}
\label{eq-09}
  D_{i\alpha,j\beta}(\tau-\tau^{\prime})
  \stackrel{\rm def}{=}
  -\langle {\rm T}_{\tau}:\phi_{i\alpha}(\tau):
  :\phi_{j\beta}(\tau^{\prime}): \rangle.
\end{equation}

\noindent Notice that the auxiliary field Green's function $D$
depends on the value of $\lambda$ introduced to make the effective
interaction matrix $\tilde{V}$ positive definite while the Green's
functions for the electrons do not. The auxiliary
phonon Green's functions are related to the electronic two-particle
response functions by the following identity, which can be derived
by integrating out the auxiliary degrees of freedom in the phonon
Green's function,

\begin{equation}
\label{eq-10}
  \chi^{-1}(k,i\omega_n)=\tilde{V}_k+ \Pi^{-1}(k,i\omega_n)
\end{equation}

\noindent with the self-energy given by the Dyson equation for the
phonons:

\begin{equation}
\label{eq-11}
  \Pi(k,i\omega_n)=-\tilde{V}^{-1}_k-D^{-1}(k,i\omega_n),
\end{equation}

\noindent where $\omega_n=(2\pi/\beta)n$ for integer $n$. 
The above two equations are
in the matrix form in the charge-spin space labeled by $\alpha=0,1,2,3$.
The bare interaction vertex $-\tilde{V}_k$ plays the role as the
free propagator and is defined as $[\tilde{V}_k]_{\alpha\beta}
\stackrel{\rm def}{=} (1/L^d)\sum_{j,l}\tilde{V}_{j\alpha,l\beta}
\exp[-ik\cdot(j-l)]$. For later use, we also write down the
Dyson equation for the electron Green's function:

\begin{equation}
\label{eq-12}
  \Sigma(k,ip_n)=ip_n-t_k-G^{-1}(k,ip_n).
\end{equation}

\noindent with $p_n=(2\pi/\beta)(n+1/2)$ for integer $n$. The
equation is in the $2\times 2$ matrix form labeled by the spin
$\sigma=\uparrow,\downarrow$.\\

With the above preparation, we now derive E-DMFT. We will consider
two situations. In the next section we describe the first case, the
E-DMFT for a homogeneous system in which all the lattice sites are
equivalent to each other. In this case E-DMFT approximation amounts
to integrating out all but one lattice site to get an effective
single site action which is equivalent to a self-consistent
impurity model. All the sites that are integrated out contribute to
self-consistent baths of free fermions and bosons for the impurity
model. Here we should have a homogeneous chemical potential
$\mu_{i\sigma}=\mu_{\sigma}$. In the second case (see \S\ref{sec-05})
we explore the E-DMFT for systems with two mutually penetrating
sublattices. We present a formalism combining E-DMFT and
C-DMFT with a cluster of two neighboring sites. While it is natural to
apply this formalism to a system with two non-equivalent sublattices,
it can also be applied to the homogeneous situation. In the latter
case, the purpose is to treat more accurately the spatial
correlations than the single site E-DMFT. This combination of
E-DMFT and C-DMFT can be formulated for clusters of arbitrary size.\\

\section{E-DMFT of a Homogeneous System}
\label{sec-03}

We first explore the single site E-DMFT applicable to a homogeneous
system. By keeping one lattice site while integrating out the rest
using the cavity construction \cite{kotliar1996}, we obtain the
E-DMFT effective action \cite{note1}

\[
  S_0^{\rm eff}=-\int_0^{\beta} d\tau \int_0^{\beta} d\tau^{\prime}
  \sum_{\sigma}
  c_{0,\sigma}^{\dagger}(\tau){\cal G}_{\sigma}^{-1}
  (\tau-\tau^{\prime}) c_{0,\sigma}(\tau^{\prime})
\]
\[
  -\frac{1}{2}\int_0^{\beta} d\tau \int_0^{\beta} d\tau^{\prime}
  \sum_{\alpha,\beta=0}^{3}
  :\phi_{0\alpha}(\tau):{\cal D}^{-1}_{\alpha\beta}(\tau-\tau^{\prime})
  :\phi_{0\beta}(\tau^{\prime}):
\]
\begin{equation}
\label{eq-13}
  +U^{\rm eff}\int_0^{\beta} d\tau :n_{0,\uparrow}(\tau):
  :n_{0,\downarrow}(\tau):
  -\int_0^{\beta} d\tau \sum_{\alpha=0}^{3} :\phi_{0\alpha}(\tau):
  :S_{0,\alpha}(\tau):
\end{equation}

\noindent This is an impurity model with both electron and  (auxiliary)
phonon degrees of freedom. The impurity model Green's functions are
identified with the local Green's functions of the lattice model, namely
[see Eqs.(\ref{eq-07})-(\ref{eq-09})],

\begin{equation}
\label{eq-14}
  G^{\rm loc}_{\sigma\sigma^{\prime}}(\tau-\tau^{\prime})
  =G_{0\sigma,0\sigma^{\prime}}(\tau-\tau^{\prime}),
  \;\;
  \chi^{\rm loc}_{\alpha\beta}(\tau-\tau^{\prime})
  =\chi_{0\alpha,0\beta}(\tau-\tau^{\prime}),
  \;\;
  D^{\rm loc}_{\alpha\beta}(\tau-\tau^{\prime})
  =D_{0\alpha,0\beta}(\tau-\tau^{\prime}).
\end{equation}

\noindent The E-DMFT self-consistent loop is formed as follows.
One starts with the effective action Eq.(\ref{eq-13}) and measures
the local electron and phonon Green's functions as define through
Eq.(\ref{eq-14}). Then one calculates the self-energies using the local
version of the Dyson equations:

\begin{equation}
\label{eq-15}
  \Sigma_{\sigma}(ip_n) =
  {\cal G}_{\sigma}^{-1}(ip_n)-[G^{\rm loc}_{\sigma}]^{-1}(ip_n),
\end{equation}
\begin{equation}
\label{eq-16}
  \Pi(i\omega_n) =
  {\cal D}^{-1}(i\omega_n)-[D^{\rm loc}]^{-1}(i\omega_n)
\end{equation}

\noindent  One of the basic assumptions of DMFT is the locality
of the self-energy \cite{kotliar1996}. Notice that from Eq.(\ref{eq-10})
the phonon self-energy is different from, although related to, the response
functions and the locality of the former does not imply the locality of the
latter. Under this assumption, the same local quantities
can be calculated, by using Eqs. (\ref{eq-10})-(\ref{eq-12}),

\begin{equation}
\label{eq-17}
  G^{\rm loc}_{\sigma}(ip_n)=\sum_{k} G_{\sigma}(k,ip_n)
  =\sum_{k} [ip_n-t_k-\mu^{\rm eff}_{\sigma}-\Sigma_{\sigma}(ip_n)]^{-1},
\end{equation}
\begin{equation}
\label{eq-18}
  \chi^{\rm loc}(i\omega_n)=\sum_{k} \chi(k,i\omega_n)
  =\sum_{k} [\tilde{V}_k+\Pi^{-1}(i\omega_n)]^{-1},
\end{equation}
\begin{equation}
\label{eq-19}
  D^{\rm loc}(i\omega_n)=\sum_{k} D(k,i\omega_n)
  =-\sum_{k}[\tilde{V}^{-1}_k+\Pi(i\omega_n)]^{-1}
\end{equation}

\noindent In the DMFT with phonons \cite{motome}, one
substitutes the results of Eqs.(\ref{eq-17}) and (\ref{eq-19}) back
in the local Dyson equations, (\ref{eq-15}) and (\ref{eq-16}), and
generates a new set of the Dynamical Weiss functions
${\cal G}$ and ${\cal D}$ which serve as the starting point
of the next iteration. This completes the self-consistent loop.\\

It is interesting to notice, however, that one has another choice to
form the self-consistent loop through the electronic two particle
Green's function $\chi^{\rm loc}$ instead of $D^{\rm loc}$, as used
in Refs.[\cite{kajueter,si2000,chitra2000}]. It is
important that the two different procedures are
compatible with each other. By combining Eqs.(\ref{eq-18}) and
(\ref{eq-19}), one can find the following relation in the matrix
form:

\begin{equation}
\label{eq-20}
  [\chi^{\rm loc}]^{-1}(i\omega_n)=
  -{\cal D}(i\omega_n)+\Pi^{-1}(i\omega_n)
\end{equation}

\noindent where the local phonon Dyson equation (\ref{eq-16}) has
been used. This is nothing but the identity one can derive directly
from the
effective action Eq.(\ref{eq-13}) and is the local version of the
electron-phonon identity, Eq.(\ref{eq-10}). Hence
the two routines, which correspond to making the DMFT approximation at
the different stages of the formulation, one directly from the non-local
electron interaction and the other after the Hubbard-Stratonovich
decomposition, are indeed equivalent, as long as the same interaction
vertex defined in Eq.(\ref{eq-04}) is used.\\

There is, though, still one point left out, that is the arbitrariness
of the value of the constants $\lambda_{\alpha\beta}$ in
Eq.(\ref{eq-04}). They were introduced in our formalism for the
purpose of ensuring the applicability of the real Hubbard-Stratonovich
transform and are thus unphysical. While it is obvious that the
artificial effects from the $\lambda$ dependences cancel out exactly
in the original action, Eq.(\ref{eq-06}), it is not clear if the same
thing happens after the E-DMFT approximation. On the other hand, the
earlier versions of E-DMFT \cite{kajueter,si2000,chitra2000} use only
the bare interactions and have no such a problem.\\

We will show in the following that, even after the E-DMFT
approximation, the effects from these $\lambda$ terms still cancel
exactly in {\it all} measurable quantities and only
the physical interactions determine the physics. In doing so, we
also establish the equivalence of our E-DMFT with the existing
ones. The physical reason behind this perfect cancellation is that
the static quartic interactions, including the $\lambda$ terms we
introduced, do not get any renormalization in DMFT, even in the
presence of those non-local interactions.\\

To proceed, we need to check the effects on the phonon dynamical
Weiss functions of the arbitrary $\lambda$ dependence we added in 
Eq.(\ref{eq-04}) for ensuring the positive-definiteness. These are
easily seen by checking the high frequency behavior of the phonon
Weiss functions. In this limit, the phonon self-energy goes to zero
at the rate of the inverse frequency square, since it is
approximately proportional to the local two particle Green's
function as shown in Eq.(\ref{eq-20}) (remember ${\cal D}$
approaches a finite constant in the same limit). So from
Eq.(\ref{eq-19}) we have:

\begin{equation}
\label{eq-21}
  D^{\rm loc}_{\alpha\beta}(i\omega_n) 
  \stackrel{n\rightarrow\infty}{\longrightarrow}
  -\sum_k \tilde{V}_{k,\alpha\beta} = -\lambda_{\alpha\beta}
\end{equation}

\noindent and thus

\begin{equation}
\label{eq-22}
  {\cal D}_{\alpha\beta}(i\omega_n)
  \stackrel{n\rightarrow\infty}{\longrightarrow}
  -\lambda_{\alpha\beta}
\end{equation}

\noindent We can make a shift of the dynamical phonon Weiss functions by
defining (in the matrix form)

\begin{equation}
\label{eq-23}
  \tilde{\cal D}(i\omega_n) \stackrel{\rm def}{=}
  {\cal D}(i\omega_n)+\lambda
\end{equation}

\noindent The Weiss function defined above approaches zero
in the high frequency limit. We then integrate out the auxiliary fields
in the effection action Eq.(\ref{eq-13}). After rearranging the variables
and introducing a new set of the auxiliary field
$\tilde{\phi}_{0,\alpha}$, the effective action becomes:

\[
  \tilde{S}_0^{\rm eff}
  =-\int_0^{\beta} d\tau \int_0^{\beta} d\tau^{\prime}
  \sum_{\sigma\sigma^{\prime}}c_{0,\sigma}^{\dagger}(\tau)
  {\cal G}_{\sigma\sigma^{\prime}}^{-1}
  (\tau-\tau^{\prime}) c_{0,\sigma^{\prime}}(\tau^{\prime})
  -\frac{1}{2}\int_0^{\beta} d\tau \int_0^{\beta} d\tau^{\prime}
  \sum_{\alpha,\beta=0}^{3}
\]
\begin{equation}
\label{eq-24}
  :\tilde{\phi}_{0\alpha}(\tau):
  \tilde{\cal D}^{-1}_{\alpha\beta}(\tau-\tau^{\prime})
  :\tilde{\phi}_{0\beta}(\tau^{\prime}):
  +U\int_0^{\beta} d\tau :n_{0,\uparrow}(\tau):
  :n_{0,\downarrow}(\tau):
  - \int_0^{\beta} d\tau \sum_{\alpha=0}^{3} 
  :\tilde{\phi}_{0\alpha}(\tau):: S_{0,\alpha}(\tau):
\end{equation}

\noindent It can be shown that, while the electronic Green's
functions remain the same, the new local phonon Green's function
$\tilde{D}^{\rm loc}$ is related to the old one by

\begin{equation}
\label{eq-25}
  \tilde{D}^{\rm loc}(i\omega_n) = D^{\rm loc}(i\omega_n)+\lambda
  +[\lambda+{\cal D}(i\omega_n)]\chi^{\rm loc}(i\omega_n)
   [\lambda+{\cal D}(i\omega_n)]
  -{\cal D}(i\omega_n)\chi^{\rm loc}(i\omega_n){\cal D}(i\omega_n)
\end{equation}

\noindent One can also find the relation between the phonon self-energies:

\begin{equation}
\label{eq-26}
  \tilde{\Pi}^{-1}(i\omega_n)=\Pi^{-1}(i\omega_n)+\lambda.
\end{equation}

\noindent The self-consistent conditions, Eqs.(\ref{eq-18}) and
(\ref{eq-19}), are transformed to

\begin{equation}
\label{eq-27}
  \tilde{\chi}^{\rm loc}(i\omega_n)
  =\sum_{k} [-V_k+\tilde{\Pi}^{-1}(i\omega_n)]^{-1},
\end{equation}
\begin{equation}
\label{eq-28}
  \tilde{D}^{\rm loc}(i\omega_n)
  =-\sum_{k}[-V^{-1}_k+\tilde{\Pi}(i\omega_n)]^{-1}.
\end{equation}

\noindent The self-consistent loop can be formed in the
same way as before.\\

So we have shown that the arbitrariness in the choice of $\lambda$ does
not affect the physical quantities. This point is best seen through
Eq.(\ref{eq-27}) where a
simultaneous shift of $V_k$ and $\tilde{\Pi}^{-1}$ cancels out exactly.
The quantities related to the auxiliary fields, including those described
in Eqs. (\ref{eq-23}), (\ref{eq-25}), (\ref{eq-26}), and (\ref{eq-28}),
do depend on $\lambda$. But as we pointed out, they are not
quantities experimentally measurable.\\

The formalism shown above, Eqs.(\ref{eq-24})-(\ref{eq-28}), is
equivalent to, although more generic than, those discussed in the
existing literatures, due to (i) the more general form of the
interaction we have taken, (ii) the consideration paid to
possible broken symmetry, and (iii) the introduction of the
$\lambda$ terms ensuring the positive-definiteness of the
interaction matrix. The method allows us to study the models with
general interactions, like the antiferromagnetic spin exchange.\\

Before we leave this part, it should be pointed out that the
formalism we just developed can readily be applied to the
general case containing electron-phonon and long range Coulomb
interactions. There is basically only one change
needed. Due to the dynamics of the real phonons from themselves,
we have an additional term $(i\omega_n)^2$ in those related
equations, including (\ref{eq-10}), (\ref{eq-11}), (\ref{eq-18}),
(\ref{eq-19}), and (\ref{eq-28}). The electron-phonon identities
still hold in
the new formalism. Without any further change, the continuous
auxiliary boson fields can be used to describe the real phonon
fields within the same formalism.\\

\section{Application I: Phase Diagram of the 3D U-V model at Half-Filling
via E-DMFT}
\label{sec-04}

\begin{figure}[h]
  \epsfxsize=12.0cm
  \centerline{\epsfbox{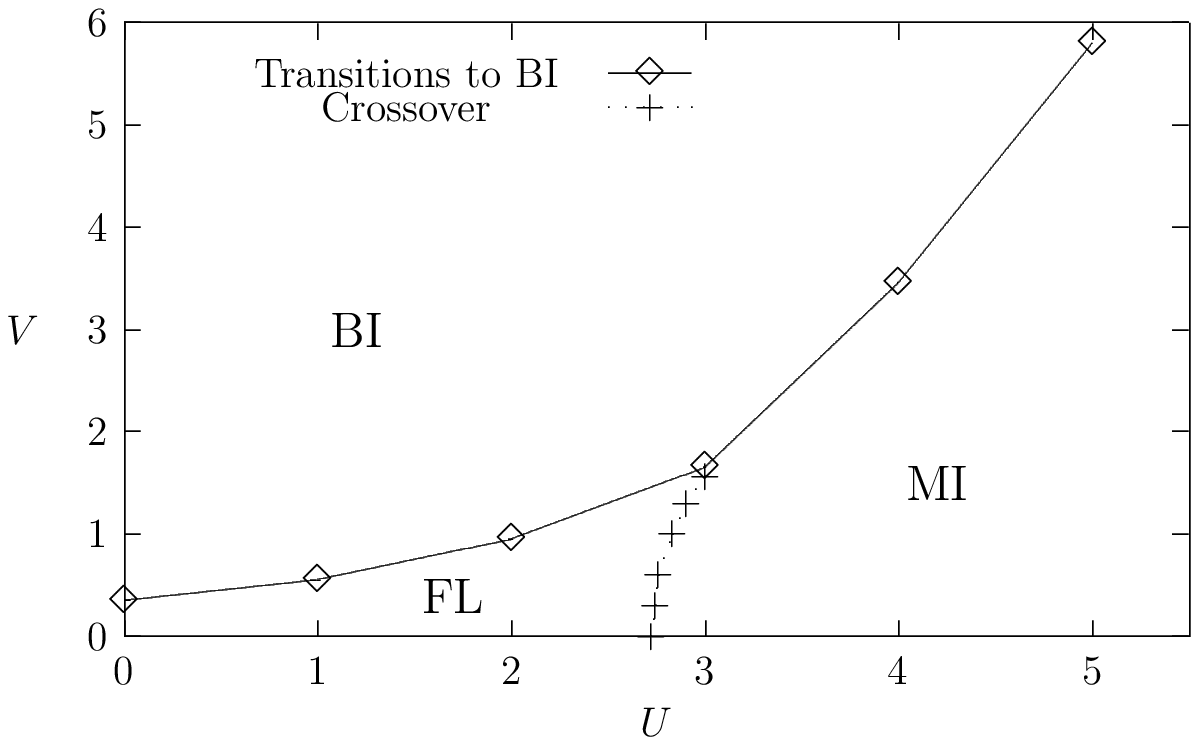}} 
  \caption{ The phase diagram of the 3D U-V model at $\beta=5.0$.
        The centers of the symbols represent the numerical
        results whose accuracy is of the order of $0.01$.
        The line bounded the BI phase represents the phase
        transitions from the FL and MI phases. The phase transition
        line is found by approaching the instability of the E-DMFT
        iteration from the smaller values of $V$. The line in
        between the FL and MI phases represents a crossover.
        It is determined by the values of $U$ and $V$ at which
        ${\rm Im}\; G_{\sigma}(ip_0)=-0.5$.}
  \label{fig-01}
\end{figure}

As an example of the practical implementation, in the following
we apply the E-DMFT formalism based on the continuous auxiliary
field approach to a 3D Hubbard model with a nearest neighbor
density-density repulsion (the U-V model). We will
be interested in the case at half-filling. This is a much
simplified version of the model we have investigated in the
last section. Especially, there is only one local shift,
$\lambda=\lambda_{00}$, needed to make the interaction matrix
positive-definite. Our purpose is to
demonstrate the implementation of E-DMFT and investigate the
behavior of the density response function in approaching the
CDW phase transition. Under the given condition, this model
Hamiltonian allows three different phases: the Mott insulating
phase when $U$ is dominant; the band insulating phase
with CDW when $V$ prevails; the metallic Fermi liquid phase
when the kinetic energy overcomes the interactions. In the MI
phase the system can develop an antiferromagnetic long order
if the magnetic frustration is weak enough. For simplicity, in the
following we will consider the system at a high enough temperature
such that the MI
phase is paramagnetic due to the strong thermal fluctuations. A
finite temperature phase diagram at $\beta=5.0$ is presented in
Fig.\ref{fig-01}. In the temperature region where our study
is performed, the Mott transition is actually a crossover
\cite{kotliar1996} so the phase diagram in Fig.\ref{fig-01}
should be viewed as a qualitative representation of the actual
phase diagram at $T=0$.\\

The 3D U-V model allows us to illustrate the novel aspects of our
methodology, including the implementation of E-DMFT with an
auxiliary field, carrying out QMC simulation with repulsive
interactions, cancellation of the $\lambda$ dependence,
and investigation of the density-density response. Physically, we
have in mind the following questions: (1) How is the MI-FL
transition affected by the non-local interaction $V$? (2) How
does the charge density instability develop in the E-DMFT
equations? (3) How does a frequency dependent effective interaction
affect the quasiparticle properties? These questions can not be
addressed in the simple DMFT studies where the $V$
interaction is handled at the Hartree level.\\

The following is the 3D model Hamiltonian we are going to study:

\[
  \hat{H}= -\frac{1}{2}\sum_{i,j}\sum_{\sigma} (t_{ij}
  \hat{c}_{i\sigma}^{\dagger}\hat{c}_{j\sigma}+\; {\rm h.c.})
  +U\sum_i :\hat{n}_{i\uparrow}: :\hat{n}_{i\downarrow}:
\]
\begin{equation}
\label{eq-29}
  +V\sum_{\langle i,j\rangle} 
  (:\hat{n}_{i\uparrow}:+:\hat{n}_{i\downarrow}:)
  (:\hat{n}_{j\uparrow}:+:\hat{n}_{j\downarrow}:),
\end{equation}

\noindent where $\langle i,j\rangle$ represents a pair of
nearest-neighboring sites. In this special case, the Fourier
transformed off-site interaction is given by

\begin{equation}
\label{eq-30}
  V_k=V[\cos(k_x)+\cos(k_y)+\cos(k_z)].
\end{equation}

\noindent We perform QMC simulation similar to that in 
Ref.[\cite{motome}]. The fermionic part of the impurity model
is handled by the standard Hirsch-Fye algorithm \cite{hirsch}.
The statistical
weight from the part of the continuous bosonic fields is
obtained directly by computing the corresponding Boltzmann
factor. We use here a semi-circular density of state for the
electronic degrees of freedom \cite{motome}. The bandwidth
is set to be the energy unit. We consider paramagnetic 
solutions at a finite temperature. Since the system is exactly at
half-filling, we use in QMC the particle-hole symmetry accompanied
by the reversal of the phonon displacement to increase the
efficiency of the simulation. In QMC, we take $\beta=5.0$ and
$\Delta\tau=1/4$. Correspondingly the inverse temperature range
$[0,\beta]$ is devided into $L=\beta/\Delta\tau=20$ slices.
The typical number of sweeps for QMC measurement
is $10^6$ by which all the quantities converge within the
statistical errors. We actually experience critical slowing
down in both QMC simulation and the DMFT self-consistent
iterations as we approach the critical point around $U=3.0$
and $V=1.6$ (see Fig.\ref{fig-01}). In this region the
typical number of iterations needed for convergence increases
from ten to twenty and we use in QMC measurement $2\sim 4\times 10^6$
sweeps.\\

\begin{figure}[h]
  \epsfxsize=14.0cm
  \centerline{\epsfbox{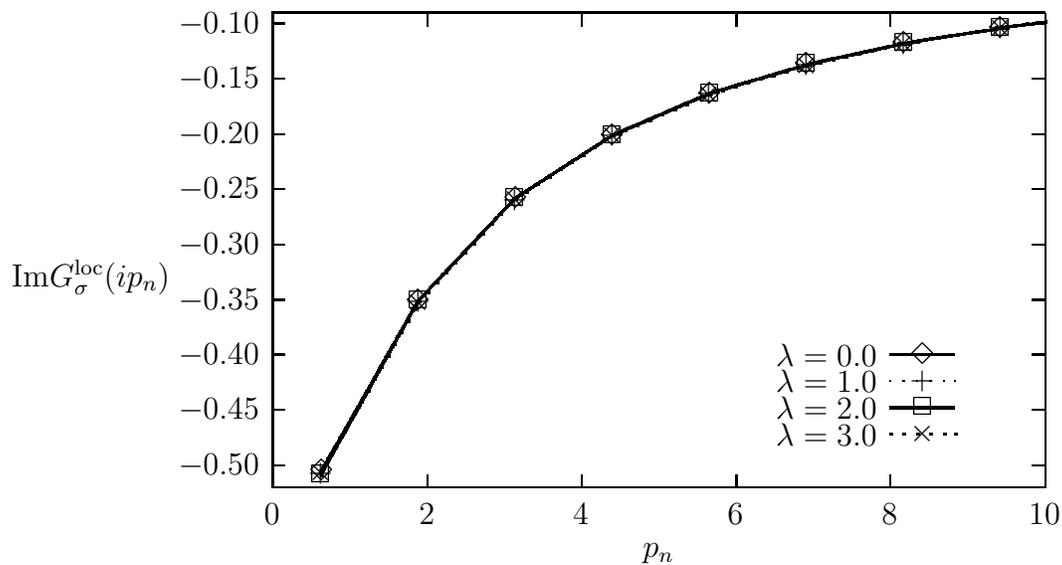}} 
  \caption{The imaginary part of the electron Green's function
    at four different values of $\lambda$ calculated at $U=3.0$
    and $V=1.6$ as a function of the Matsubara frequency \cite{note2}.}
  \label{fig-02}
\end{figure}

\begin{figure}[h]
  \epsfxsize=14.0cm
  \centerline{\epsfbox{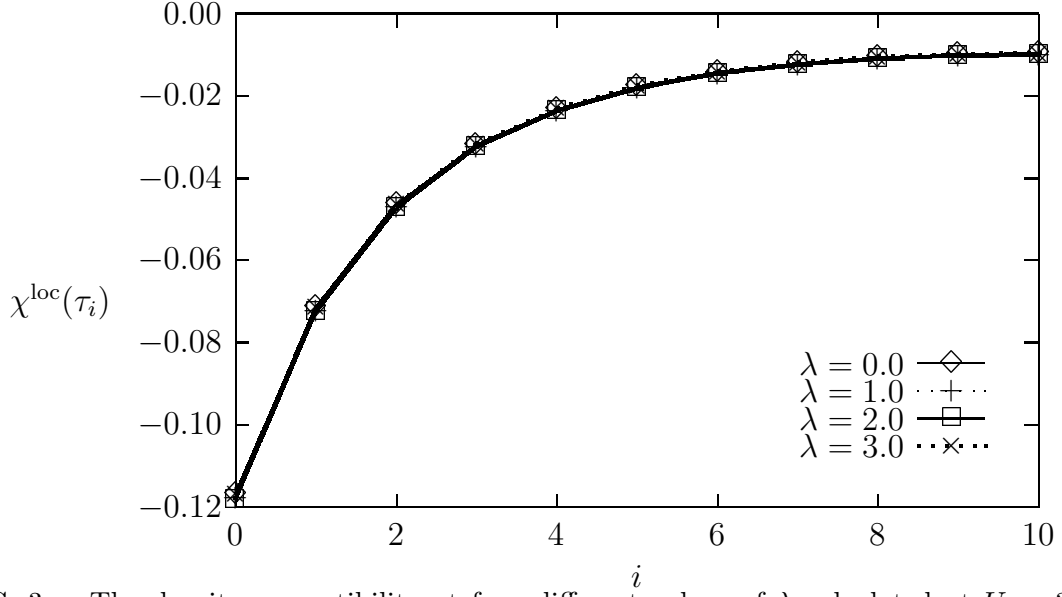}} 
  \caption{ The density susceptibility at four different values of 
    $\lambda$ calculated at $U=3.0$ and $V=1.6$ as a function of
    the imaginary time.}
  \label{fig-03}
\end{figure}

Before we present the results, we want to show that at the numerical
level, the $\lambda$ depending term does not affect the physical
results. We have shown in the previous section that in the
formulation of E-DMFT the $\lambda$ dependence does not show up in
physical quantities. However it can happen that if the formalism 
is so sensitive to the dependence such that in a practical calculation
like QMC there is always only partial cancellation and a significant
$\lambda$
dependence remains in the physical measurables. It is also possible
that a negligible $\lambda$ dependence in the results of the impurity
model, which is always there unless one can solve the problem exactly,
may get magnified during the E-DMFT iterations. In the following we
present the results calculated at $U=3.0$ and $V=1.6$ which show no
$\lambda$ dependence. From the phase diagram
Fig.\ref{fig-01} one can see that this is the point we have
reached closest in the parameter space to the finite temperature
critical point (CP). If the suggested scenarios of the $\lambda$
dependence in the physical results may happen, they most likely
happen around the CP where the system becomes very sensitive to the
extra change from $\lambda$. In Fig.\ref{fig-02} we
plot the imaginary part of the electron Green's functions calculated
at four different $\lambda$'s. The result shows no $\lambda$
dependence within the accuracy of the
calculation \cite{note2}. Similar for the electron density Green's
functions as shown in Fig.\ref{fig-03}. So in both the
formulation and the practical calculation, the $\lambda$ term does not
play any physical role and we can use it safely. In all the following
calculation, we set $\lambda=2.0$ in the QMC simulation.\\

\subsection{Near the Mott Crossover Line}

\begin{figure}[h]
  \epsfxsize=14.0cm
  \centerline{\epsfbox{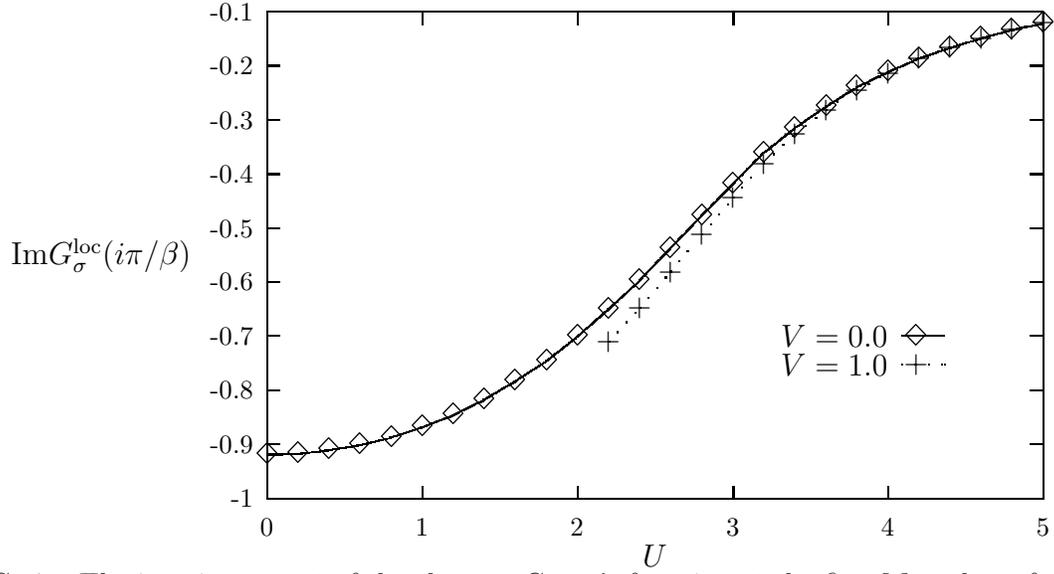}}
  \caption{ 
        The imaginary part of the electron Green's function at
        the first Matsubara frequency,
        ${\rm Im}\; G^{\rm loc}_{\sigma}(i\pi/\beta)$,
        as a function of the on-site interaction $U$ for two
        different values of $V$.}
\label{fig-04}
\end{figure}

We first present the result showing the crossover between the MI and
FL phases. In Fig.\ref{fig-04} we show the data of
${\rm Im}\; G^{\rm loc}_{\sigma}(i\pi/\beta)$
v.s. $U$ at $V=0.0$ and $1.0$. The reason
we choose to plot this function is that it is known \cite{kotliar1996}
that the asymptotic behavior at the low frequency limit of the imaginary
part of the electron Green's function reflects the density of states
(DOS) near the Fermi surface. For the metallic phase the DOS is finite
at the Fermi surface and hence
${\rm Im}\; G(ip_n)\stackrel{p_n\rightarrow 0}{\longrightarrow}
{\rm Const.} \neq 0$. For the insulating phase, on the other
hand, the corresponding DOS is zero and
${\rm Im}\; G(ip_n)\stackrel{p_n\rightarrow 0}{\longrightarrow} 0$.
So from the plotting of ${\rm Im}\; G$ at the first fermionic
Matsubara frequency we can study the transition or crossover behavior
between the two phases. Since we solve the problem at a temperature
($\beta=5.0$) higher than the critical temperature ($\beta_c\sim 1/0.04$,
see Ref.[\cite{kotliar1996}]) of the
Mott transition, what we see in Fig.\ref{fig-04} is that for fixed
$V$, $-{\rm Im}\; G^{\rm loc}_{\sigma}(i\pi/\beta)$
decreases smoothly as $U$ is increased.
This behavior represents the crossover between the two phase. There
is a small but finite shift in between the two curves
calculated at $V=0.0$ and $1.0$. This is a result of the competition
between $U$ and $V$. In the case of $V=1.0$ the system
enters the BI phase for $U\alt 2.2$ which one can see approximately from
Fig.\ref{fig-01}. This region is out of the scope of the current
approach applicable only to a homogeneous system. In the range
$2.2\alt U \alt 4.0$,
which is, roughly speaking, the crossover region, the value of
${\rm Im}\; G^{\rm loc}_{\sigma}(i\pi/\beta)$
at the same $U$ is always bigger for $V=1.0$ than
that for $V=0.0$. This means that the existence of the finite $V$ makes
the system more metallic and so the effective $U$ is smaller, as one
anticipates from qualitative considerations. If $U$ is increased
further so that $4.0\alt U$, both the systems enter the paramagnetic
insulating phase with literally no difference.\\

\begin{figure}[h]
  \epsfxsize=14.0cm
  \centerline{\epsfbox{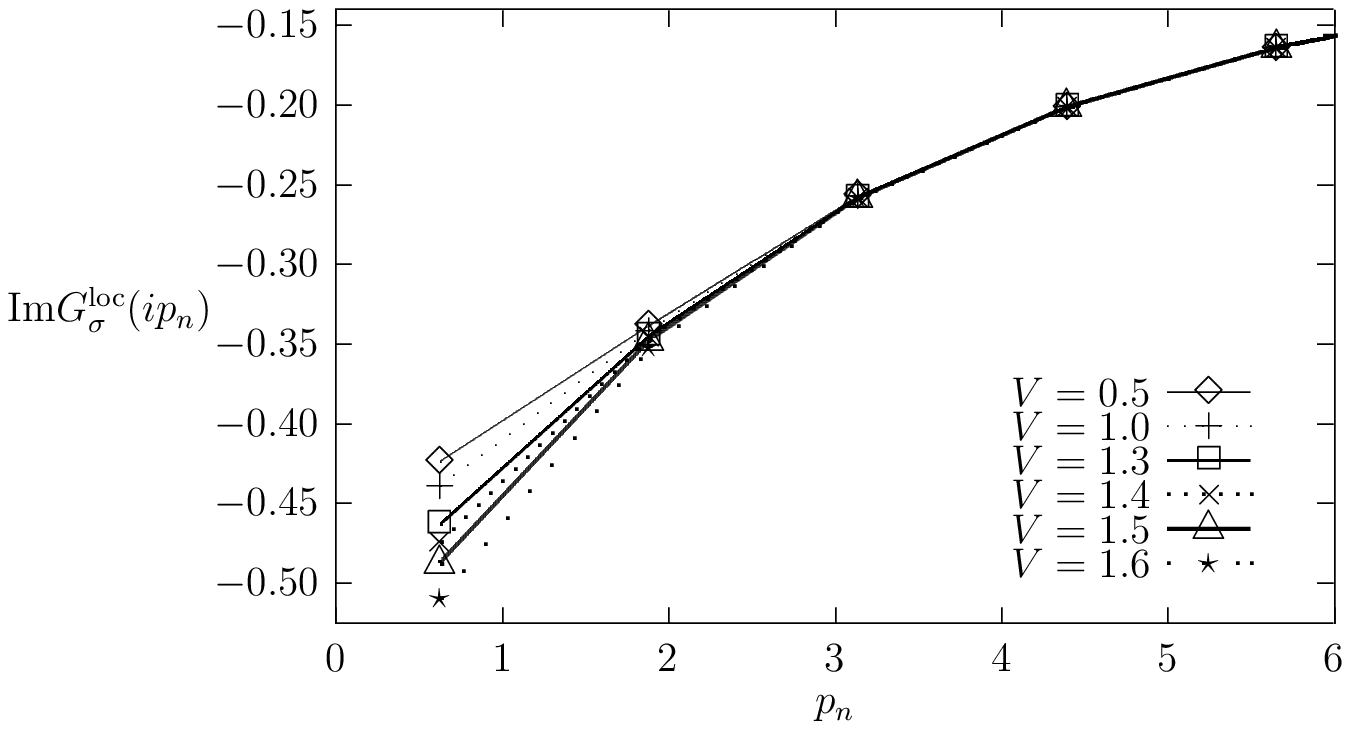}}
  \caption{ 
        The imaginary part of the local electron Matsubara Green's
        function for different values
        of the interaction $V$ at fixed $U=3.0$.}
\label{fig-05}
\end{figure}

\begin{figure}[h]
  \epsfxsize=14.0cm
  \centerline{\epsfbox{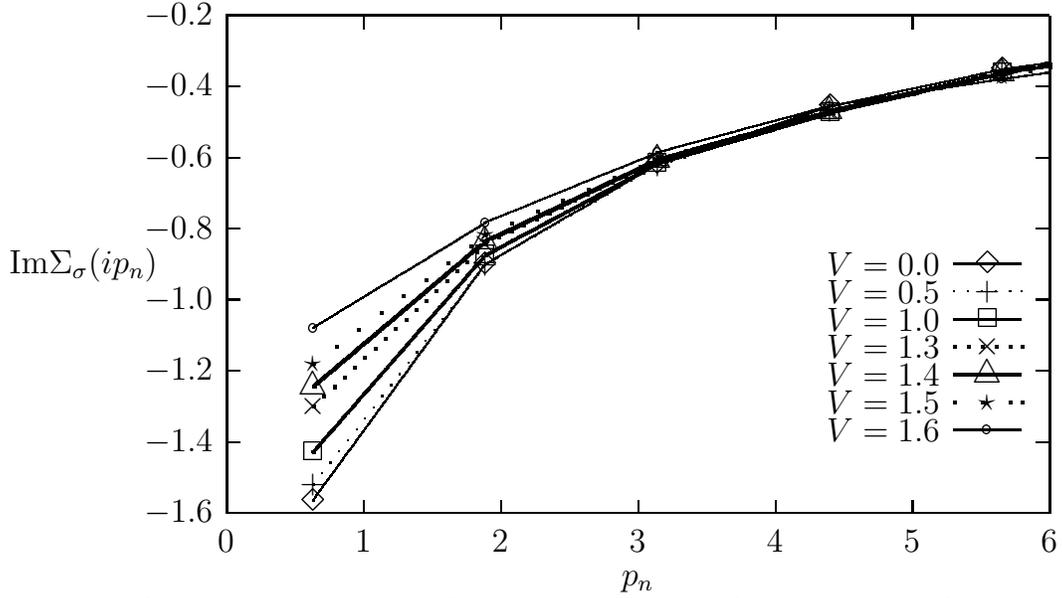}}
  \caption{ 
        The frequency dependence of the imaginary part of the electron
        self-energy $\Sigma_{\sigma}(ip_n)$ at different values
        of the interaction $V$ at fixed $U=3.0$.}
\label{fig-06}
\end{figure}

\begin{figure}[h]
  \epsfxsize=14.0cm
  \centerline{\epsfbox{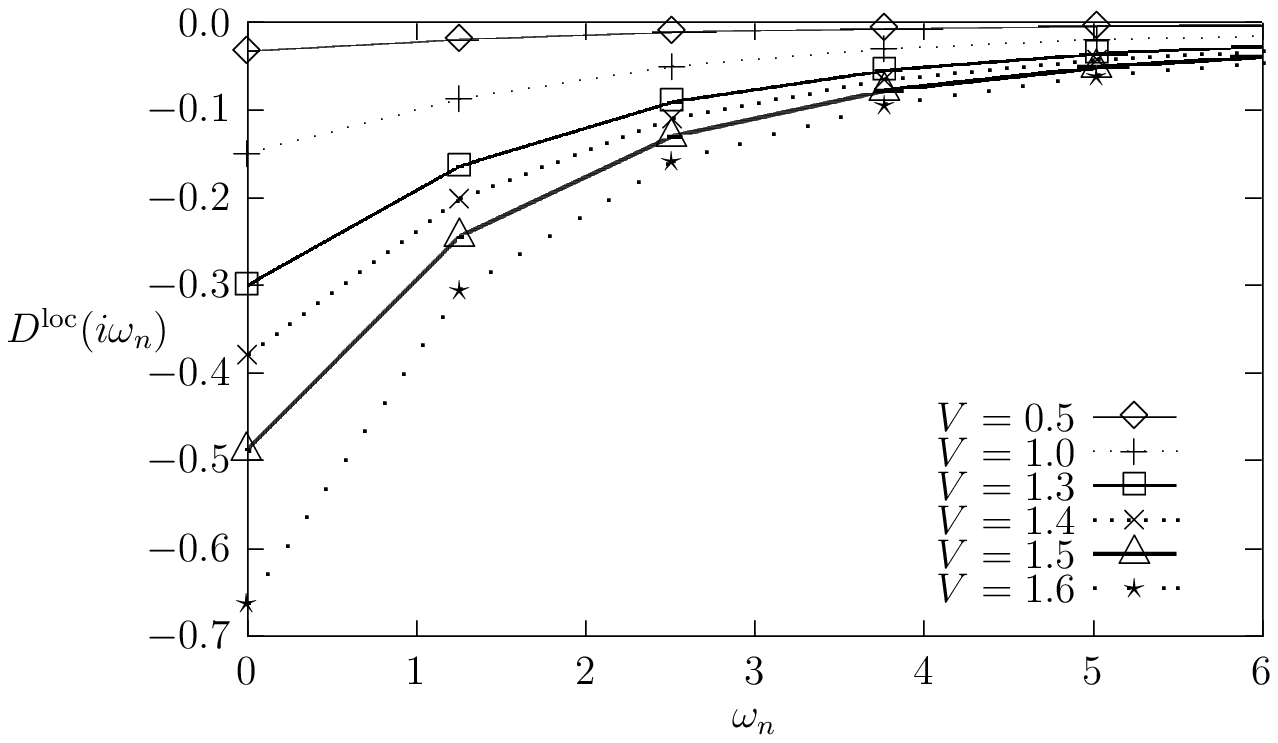}}
  \caption{The local phonon Matsubara Green's function for different
        values of the interaction $V$ at fixed $U=3.0$ \cite{note3}.}
\label{fig-07}
\end{figure}

We then investigate the behavior of the various Green's functions as
the transition towards the band insulating phase is approached by
changing $V$ at fixed $U$. We first show the results at $U=3.0$.
We solve the problem for seven different values of the
interaction: $V=0.5$, $1.0$, $1.3$, $1.4$, $1.5$, $1.6$, and $1.7$.
In all the results presented below the data from the last case are
not shown because we already encounter the instability at which the
convergence of the self-consistent interaction is lost.
So we can view $V=1.7$ as an upper boundary of the metallic
phase for $U=3.0$. In Fig.\ref{fig-05} we show the imaginary
part of the electron Green's function and in Fig.\ref{fig-06}
the imaginary part of the electron self-energy. It can be seen
clearly that, while the trends of the change of the plotted quantities
are in the direction towards a more metallic phase as $V$ is increased
(that is, bigger ${\rm Im} G$ and smaller self-energy at the first
several Matsubara frequencies), the magnitudes of the
changes are very limited, especially in comparison with the
phonon Green's functions \cite{note3} shown in Fig.\ref{fig-07}.
The local phonon
Green's function, which is related to the density susceptibility
as we have shown in the last section, increases significantly
as we approach the phase transition point. We can analyze the behavior
from the self-consistent equation (\ref{eq-19}), which can be
rewritten as

\begin{equation}
\label{eq-31}
  D^{\rm loc}(i\omega_n)
  =\sum_{k}\frac{V_k}{{1-V_k\Pi}(i\omega_n)}.
\end{equation}

\begin{figure}[h]
  \epsfxsize=14.0cm
  \centerline{\epsfbox{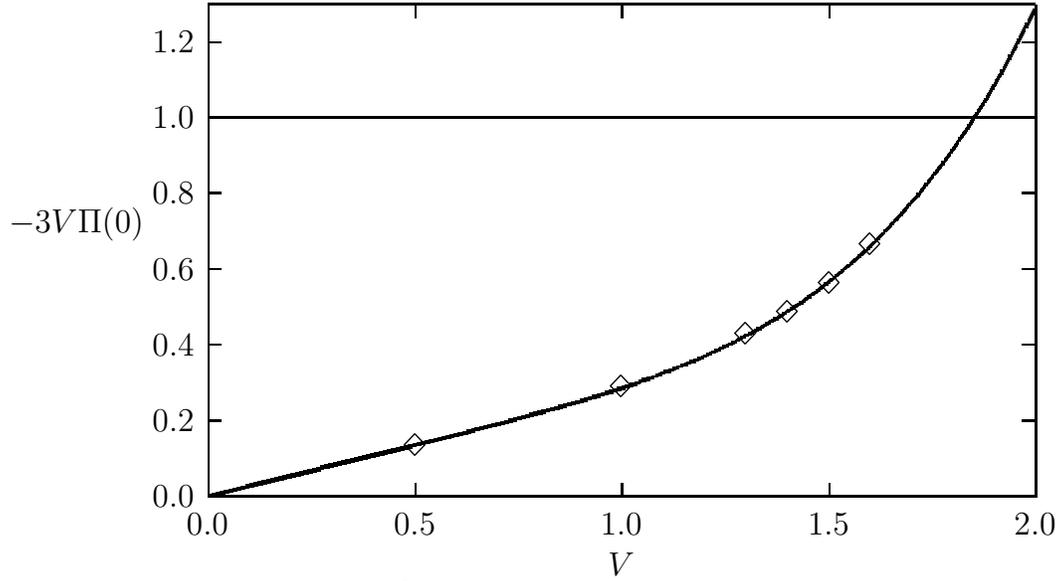}}
  \caption{ 
        The rescaled phonon self-energy at the most instable point
        with $k=(\pi,\pi,\pi)$ and $\omega_n=0$ for $U=3.0$. The
 	diamonds represent the calculated data. The curve is plotted
	 using a polynomial fitting of the phonon self-energy $\Pi$
	(which is even in $V$) up to
        $V^4$. From the extrapolation the transition is at $V_c\simeq 1.85$.
        This value is less accurate than the one from QMC due to
        the non-analyticity near the transition.
   }
\label{fig-08}
\end{figure}

\noindent As a result of screening the phonon self-energy is negative.
Moreover, $|\Pi(i\omega_n)|$ is a monotonically decreasing
function of the frequency, since the screening becomes less effective
as the frequency increases. Hence the instability, if it may happen,
will first show up at the wave vector
$k=q\stackrel{\rm def}{=}(\pi,\pi,\pi)$ and the frequency
$\omega_n=\omega_0=0$,
where the product
$V_q\tilde{\Pi}(i\omega_0)=-3V\tilde{\Pi}(i\omega_0)$ has
the biggest positive value. In Fig.\ref{fig-08} we show the plotting
of $-3V\tilde{\Pi}(i\omega_0)$ v.s. $V$ at the given 
$U$. The trend is obvious for the
product to approach ``$1$'' where the corresponding denominator in
Eq.(\ref{eq-31}) vanishes. This establishes the picture that as the
transition point is approached, the denominator disappears first at
$q$ and $\omega_0$, which corresponds to an instability against the
homogeneous ground state with a static CDW at wave vector $(\pi,\pi,\pi)$.
This is a typical phenomenon in the Green's function description of
phase transitions \cite{doniach}, although the quantities involved
here are non-perturbative. The instability is signaled by a frequency
of phonon becoming negative \cite{motome,pankov}. At this point an
ordered mean field state would be the correct solution. It should be
noted that, even when the instability happens, one may still continue
the paramagnetic solution of E-DMFT by taking the principle part in
Eq.(\ref{eq-31}). But in 3D the convergence near transition is
impossible because the derivative of the phonon Green's
function with respect to the control parameter $V$ becomes infinite.\\

\subsection{Frequency Dependent On-Site Interaction}

\begin{figure}[h]
  \epsfxsize=14.0cm
  \centerline{\epsfbox{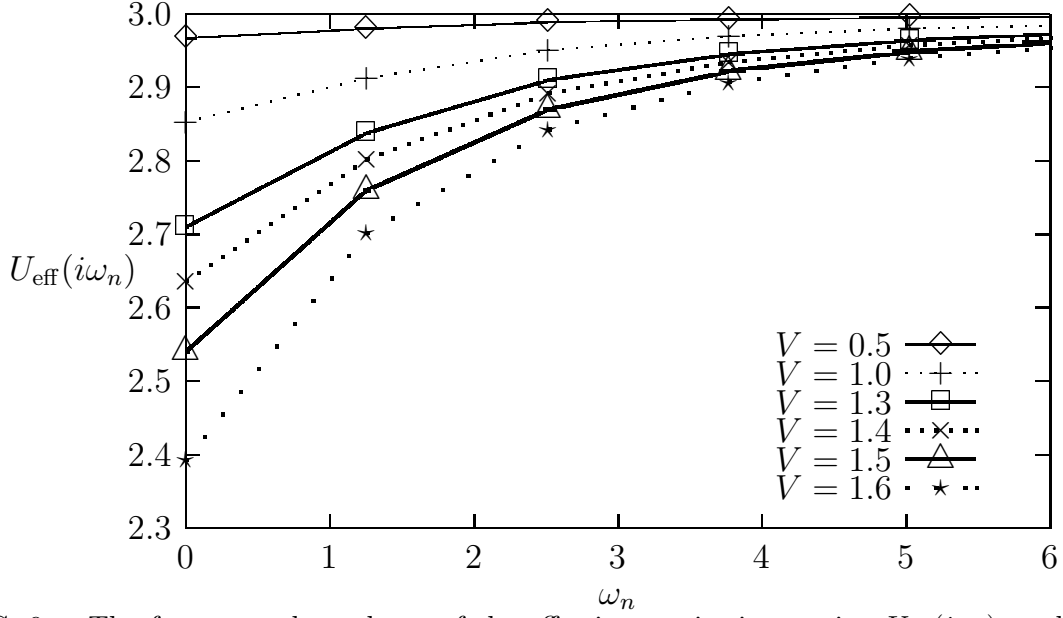}}
  \caption{ 
        The frequency dependence of the effective on-site
        interaction $U_{\rm eff}(i\omega_n)$ at different values
        of the interaction $V$ at fixed $U=3.0$.}
\label{fig-09}
\end{figure}

\begin{figure}[h]
  \epsfxsize=14.0cm
  \centerline{\epsfbox{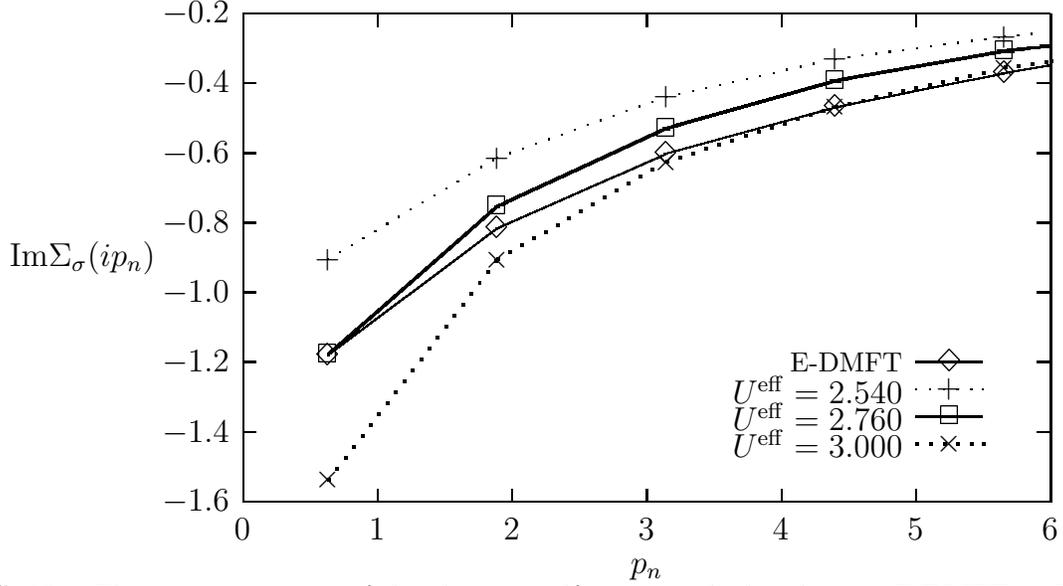}}
  \caption{ The imaginary part of the electron self-energy calculated
    using E-DMFT at $U=3.0$ and $V=1.5$ and at the corresponding
    $U^{\rm eff}=U_{\rm eff}(i\omega_n)$ for $n=0,1,\infty$ with
    $V=0$. We used $U_{\rm eff}(i\omega_0)=2.540$ and
    $U_{\rm eff}(i\omega_1)=2.760$ from our E-DMFT calculation.
    Obviously $U_{\rm eff}(i\omega_{\infty})=3.0$
   }
\label{fig-10}
\end{figure}

Another interesting property to investigate is the effective
on-site density-density interaction, which is defined at $\lambda=0$

\[
  U_{\rm eff}(i\omega_n) \stackrel{\rm def}{=}
  U+{\cal D}(i\omega_n)
\]

\noindent and is frequency dependent. From Fig.\ref{fig-09} we
can see clearly that as the transition is approached, there is a
tendency of the softening of the effective interaction at the
zero frequency. In our model the frequency dependence of the
local $U_{\rm eff}$ is due to screening of the bare interaction
by the {\it intersite} Coulomb interaction $V$. Notice however
that a frequency dependent $U$ occurs more generally in realistic
models of correlated electron due to {\it intrasite} screening
by other local orbitals as a recent local GW calculation \cite{zein}
shows. This is the first E-DMFT study of a model where the
interaction $U$ is frequency dependent. We should stress that this
behavior of the single particle Green's function can not be
described by an ordinary DMFT with fixed $U$. In Fig.\ref{fig-10}
we show how the frequency depending effective $U$ changes the
single electron behavior. We plot in Fig.\ref{fig-10} the imaginary
part of the
electron self-energy as a function of the Matsubara frequency. At
high frequencies, the self-energy from E-DMFT coincides with
that calculated using the bare $U$ alone. This tells that the screening
effect is not effective in the high frequency limit, same as
we can see from the effective $U$ plotted in Fig.\ref{fig-09}.
In the low frequencies, the
self-energy deviates to that of a smaller effective $U$. In
Fig.\ref{fig-10} we plotted the self-energies calculated at $V=0$
and $U^{\rm eff}=U_{\rm eff}(i\omega_n)$ evaluated at the lowest and
the next lowest Matsubara frequencies. We can see that
at the first Matsubara frequency the E-DMFT self-energy is closest to
that given by $U_{\rm eff}(i\omega_1)$ and $V=0$.
We can understand the situation
by thinking that there are two different $U$'s controlling low and high
frequency regions separately. Some effective and screened $U$ is in
charge of the low
frequency behavior while the bare $U$ works in the high frequency
region. In between there is a kind of the crossover connecting the
two. The results shown here suggests that a frequency independent
effective $U$ is not enough to capture the physics in the entire
frequency range.\\

The set of diagrams presented above are plotted very close to the line of
the Mott crossover (see Fig.\ref{fig-01}). In the following, we show
two other sets of data
which are plotted in the metallic and the Mott insulator phases, 
respectively.\\

\subsection{Metallic Phase}

\begin{figure}[h]
  \epsfxsize=14.0cm
  \centerline{\epsfbox{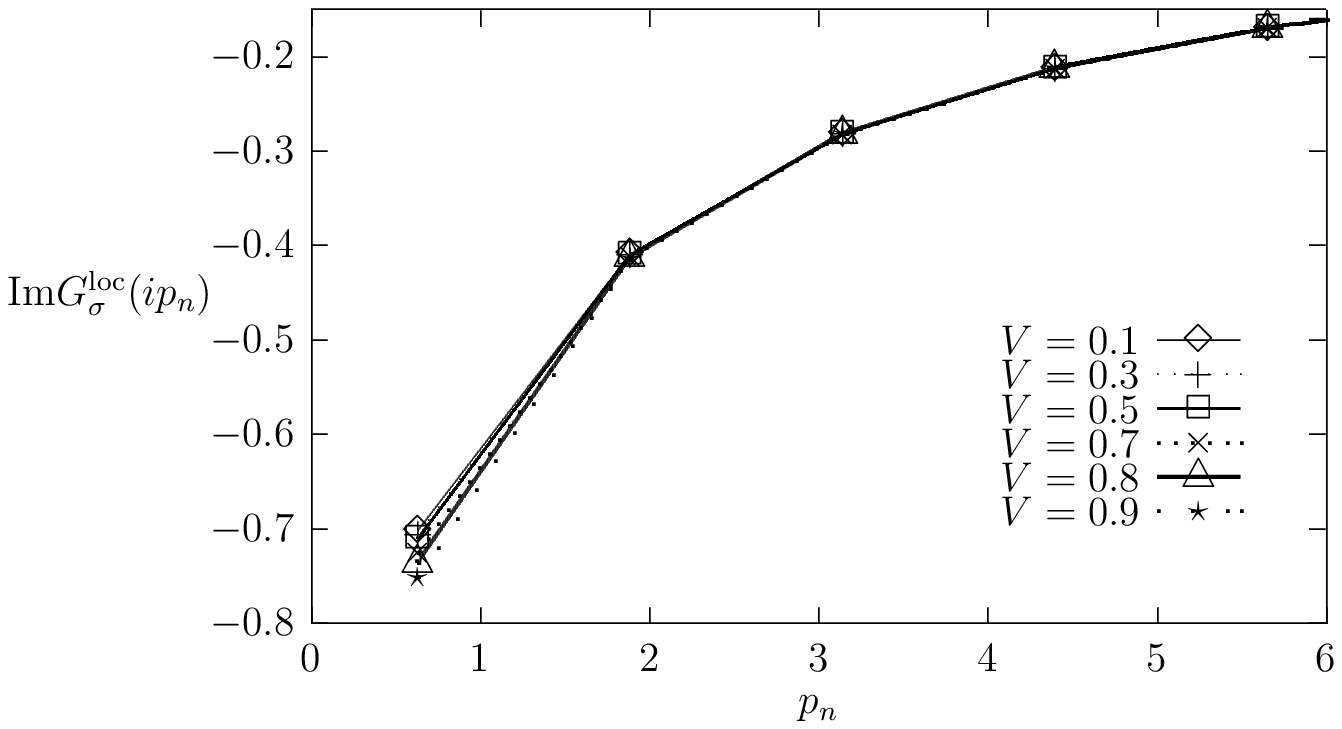}}
  \caption{ 
        The imaginary part of the local electron Matsubara Green's
        function for different values
        of the interaction $V$ at fixed $U=2.0$.}
\label{fig-11}
\end{figure}

\begin{figure}[h]
  \epsfxsize=14.0cm
  \centerline{\epsfbox{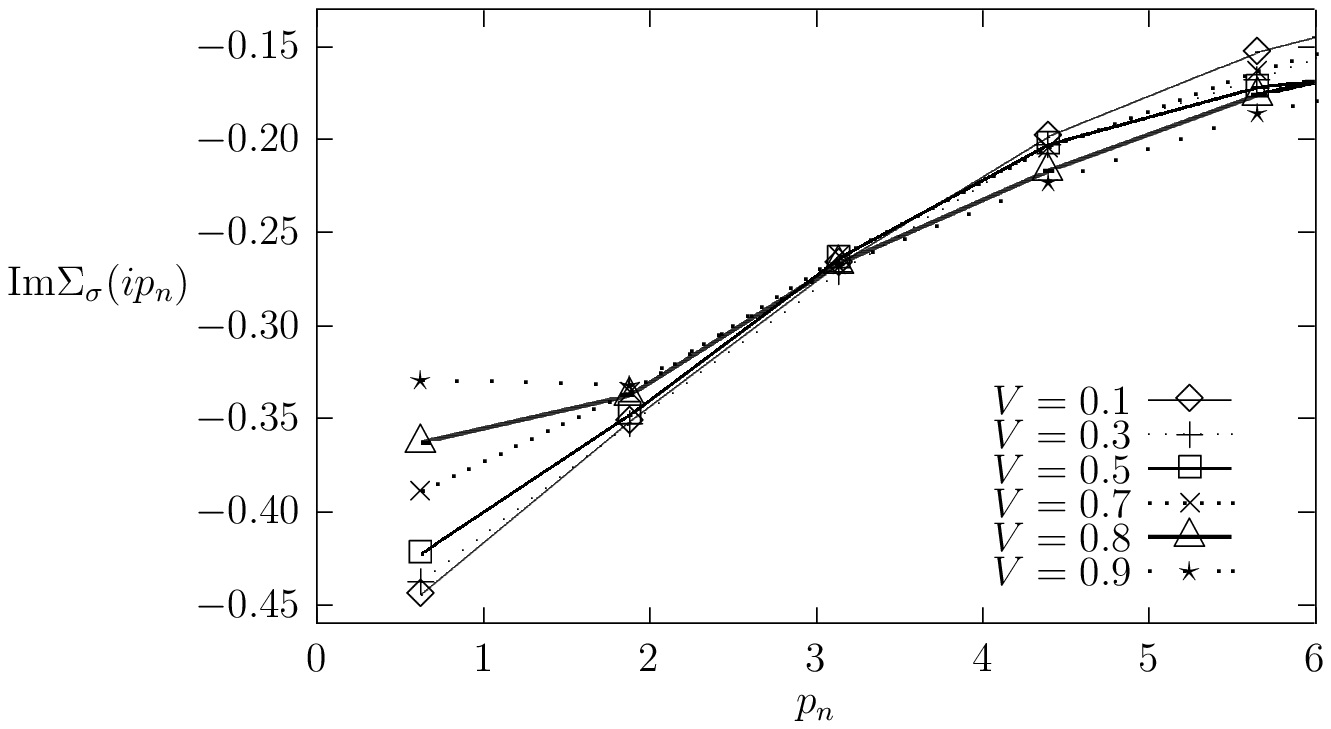}}
  \caption{The imaginary part of the local electron self-energy for
    different values
    of the interaction $V$ at fixed $U=2.0$.}
\label{fig-12}
\end{figure}

\begin{figure}[h]
  \epsfxsize=14.0cm
  \centerline{\epsfbox{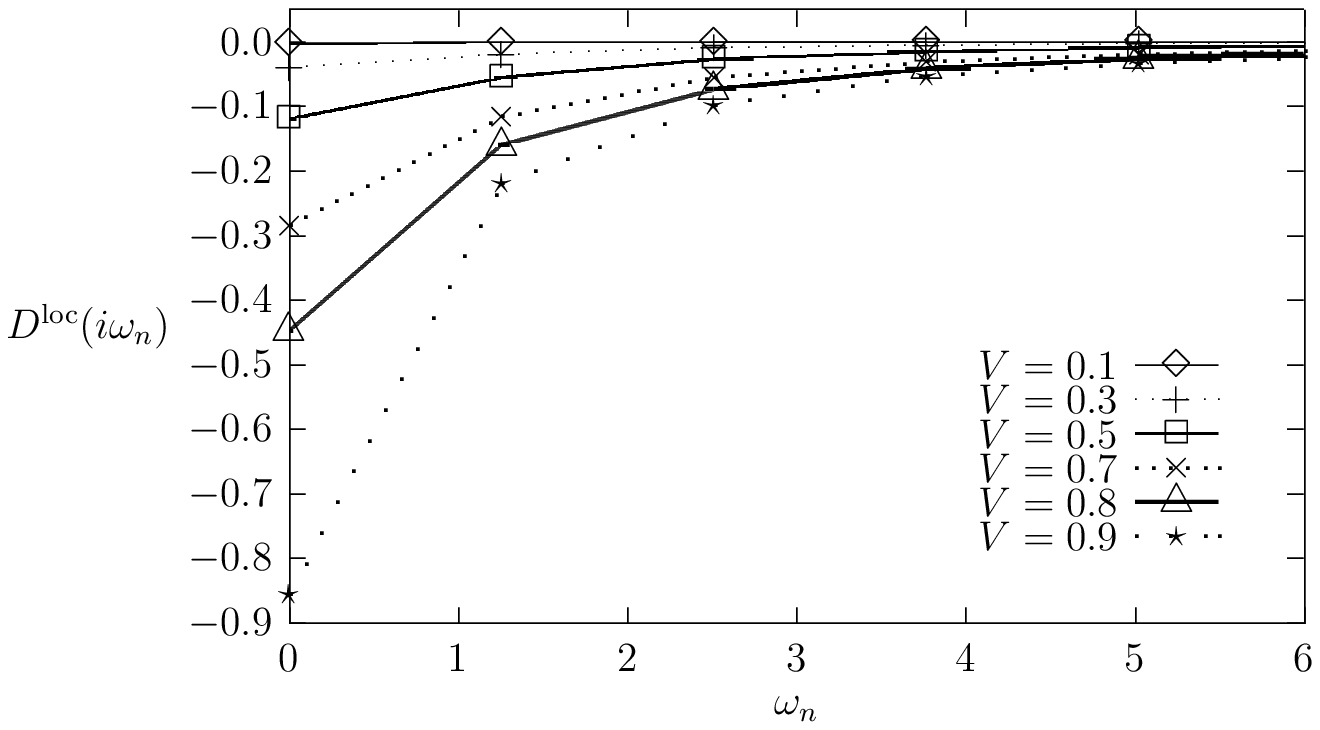}}
  \caption{ 
        The local phonon Matsubara Green's function for different values
        of the interaction $V$ at fixed $U=2.0$.}
\label{fig-13}
\end{figure}

First, we show the results at $U=2.0$ and increasing $V$ in
Figs.\ref{fig-11}-\ref{fig-13}.  As can be seen from
Fig.\ref{fig-01}, at this set of parameters  the system is in
the correlated metallic phase. To be concise, we show here three
representative plottings, that is the electron and
phonon Green's functions and the electron self-energy. One can see
that in this phase the change of the single electron Green's function
is very limited as the transition is approached at $V\sim 0.95$.
Meanwhile, the single electron self-energy changes quite a lot in
the low frequency region, showing
the stronger cancellation between the effects from $U$ and $V$ and thus
the more significant reduction of the self-energy as $V$ is increased.
The change of the phonon Green's function is again much bigger than those
of the electrons.\\

\subsection{Paramagnetic Mott Insulating Phase}

\begin{figure}[h]
  \epsfxsize=14.0cm
  \centerline{\epsfbox{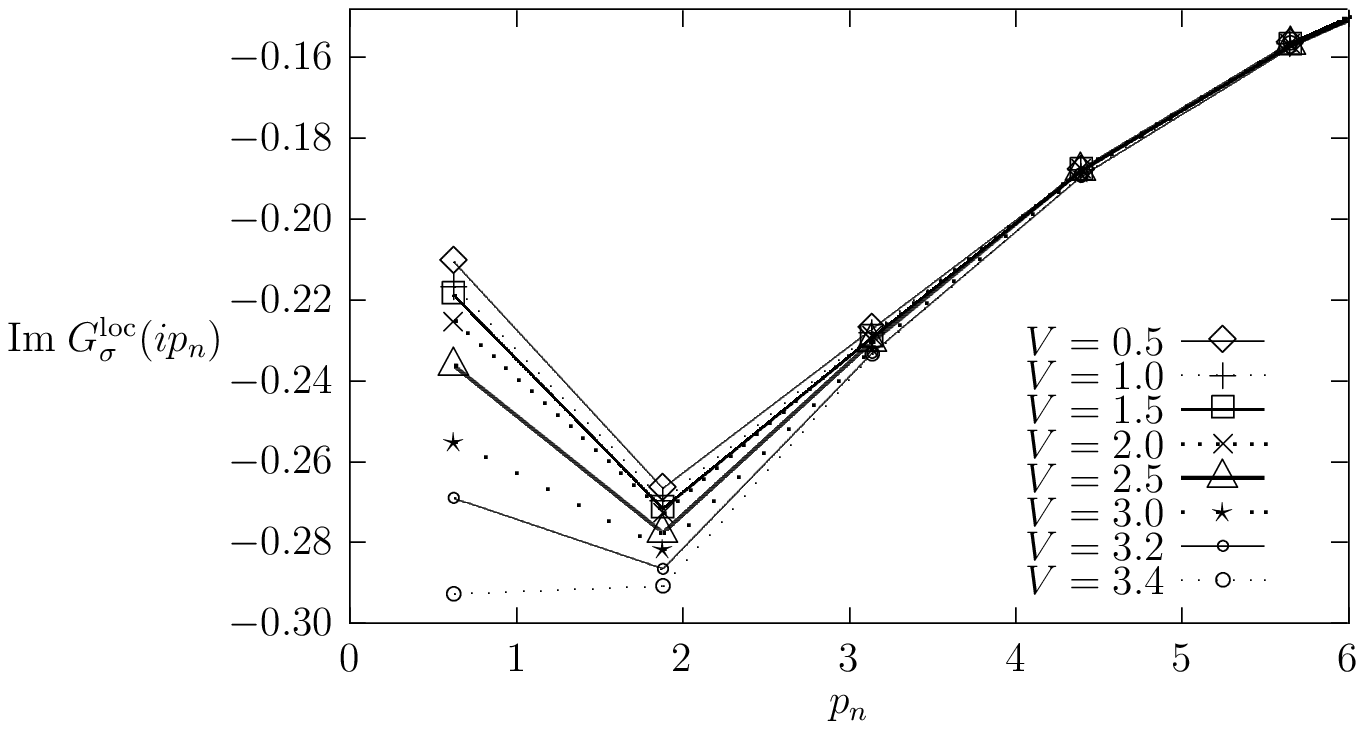}}
  \caption{ 
        The imaginary part of the local electron Matsubara Green's
        function for different values
        of the interaction $V$ at fixed $U=4.0$.}
\label{fig-14}
\end{figure}

\begin{figure}[h]
  \epsfxsize=14.0cm
  \centerline{\epsfbox{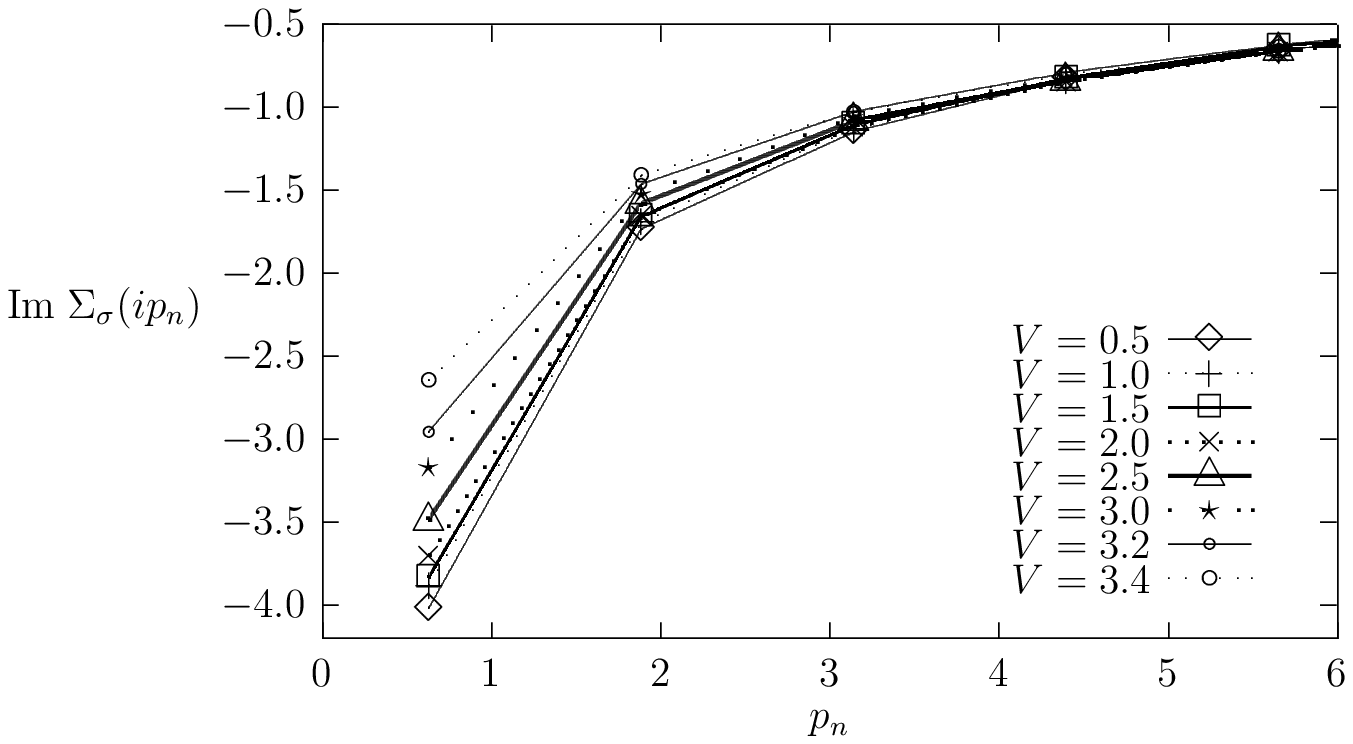}}
  \caption{The imaginary part of the local electron self-energy for
    different values
    of the interaction $V$ at fixed $U=4.0$.}
\label{fig-15}
\end{figure}

\begin{figure}[h]
  \epsfxsize=14.0cm
  \centerline{\epsfbox{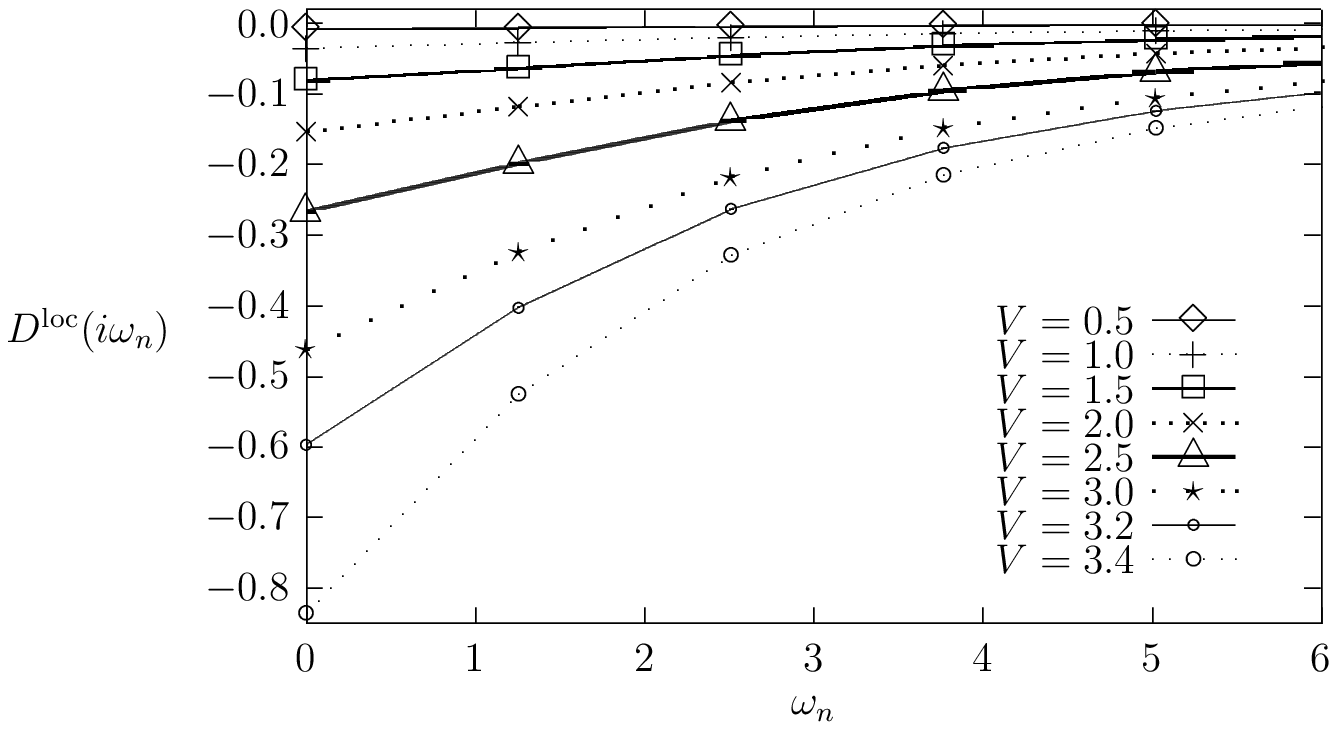}}
  \caption{ The local phonon Matsubara Green's function for different values
        of the interaction $V$ at fixed $U=4.0$.}
\label{fig-16}
\end{figure}

The second case with $U=4.0$ is shown in 
Figs.\ref{fig-14}-\ref{fig-16}. Here we work in the Mott insulating
phase. One can see from the asymptotic behavior of the Green's functions in
the low frequency limit that in entire the range, especially near the
transition at $V_c\sim 3.4$, the system is still in the
Mott insulating phase. Meanwhile, the corresponding phonon Green's functions
plotted in Fig.\ref{fig-16} change a lot.\\

\subsection{Phase Diagram}

Finally, by literally sweeping across the U-V space, we are able to
establish the
finite temperature phase diagram presented in Fig.\ref{fig-01}.
The phase transition from the metallic and the MI phases to the
BI phase is determined unambiguously from the breaking down of the
convergence of the E-DMFT iterations. We locate the crossover line between
the FL and MI phases by search the points of (U,V) at which
${\rm Im} G(ip_0)=-0.5$. While its specific value is arguable,
this criterion works well practically in the sense that right
around the CP it suggests, $(U_c,V_c)\sim (3.0,1.6)$,
we experienced the strongest critical slowing down.\\

Two remarks are in place concerning the qualitative features
contained in this finite temperature phase diagram.
First, the slopes of the boundaries of the
FL phase are positive on both sides.
This actually reflects the competition between $U$ and $V$:
The existence of a finite and small $V$ requires a bigger $U$ in
order to make the Mott transition or crossover.
Similarly a finite and small $U$ makes it harder to develop the CDW.
Second, the effects of a finite $V$ is much stronger than that of $U$,
because of the coordination number, which is six in the current
case. The above features will retain in the phase diagram at $T=0$.\\

\section{E-DMFT plus C-DMFT for Systems with Two Sublattices}
\label{sec-05}

Next we consider E-DMFT on systems with two interpenetrating sublattices.
The current study is useful in
the situation when the two sublattices are not equivalent in the
sense that, while it is homogeneous within each of them, the
order parameter is different in the two sublattices. One then
needs in E-DMFT a cluster containing at least two neighboring
sites. It is interesting, though,
to notice that the formalism we are going to develop also applies
to the homogeneous systems. In this case the cluster plays the role
to improve the description of the spatial correlations.
The formalism described in this section
can be easily extended to problems where clusters of bigger sizes
are needed. E-DMFT was combined with Dynamical Cluster Approximation (DCA)
in Ref.[\cite{jarrell}].\\

Under the given conditions, the nearest neighbor hoppings and
interactions alway connect the two different sublattices, the next
nearest neighbor ones are within the same sublattice, {\it etc}.
We rely on the external magnetic fields introduced at the beginning,
Eq.(\ref{eq-01}), to lift any possible degeneracy in the ground state.
To illustrate the basic idea while avoid any unnecessary repetition
(as it will turn out, the E-DMFT with two sublattices shares many
properties with that for the homogeneous system) , we work on
a model with only nearest neighbor hoppings and density-density
interactions, besides the on-site energy (the chemical potential)
and the Hubbard interaction. The Hamiltonian reads:

\[
  \hat{H}= -t \sum_{\langle Ai,Bj \rangle \sigma} (
  \hat{c}_{Ai,\sigma}^{\dagger}\hat{c}_{Bj,\sigma}+\; {\rm h.c.})
  -\sum_{Xi,\sigma} \mu_{Xi,\sigma} \hat{n}_{Xi,\sigma}
\]
\begin{equation}
\label{eq-32}
  +U\sum_{Xi} \hat{n}_{Xi,\uparrow} \hat{n}_{Xi,\downarrow}
  + \sum_{\langle Ai,Bj \rangle} 
  (\hat{n}_{Ai,\uparrow}+\hat{n}_{Ai,\downarrow}) V_{Ai,Bj}
  (\hat{n}_{Bj,\uparrow}+\hat{n}_{Bj,\downarrow}),
\end{equation}

\noindent where every site is label by $Xi$ with $X$ labeling
the two sublattices, $X=A,B$, and $i$ the coordinate within the sublattice.
$\langle Ai,Bj \rangle$ represents a pair of neighboring
sites. We choose the chemical potential consistent with the two
sublattice picture:

\begin{equation}
\label{eq-33}
   \mu_{Xi,\sigma}=\left\{ \begin{array}{cr}
                              \mu_{\rm A\sigma} \;\;\;\;\;& X= A \\
                              \mu_{\rm B\sigma} \;\;\;\;\;& X= B \\
                 \end{array}
         \right.
\end{equation}

\noindent  We set for nearest neighbors $V_{\langle Ai,Bj \rangle}=V \neq 0$.
We then introduce the $\lambda$ term same as before

\[
  \tilde{V}_{Xi,Yj}=\lambda\delta_{XY}\delta_{ij}-V_{Xi,Yj}
\]

\noindent with $\lambda$ a constant which ensures the $\tilde{V}_{2\times 2}$ 
matrix to be positive-definite. After normal ordering the operators in
the interactions and perform the Hubbard-Stratonovich transform, we have
the effective action:

\[
   S= \int_0^{\beta} d\tau \{ \sum_{Xi,\sigma} [
   c_{Xi,\sigma}^{\dagger}(\tau) \partial_{\tau} c_{Xi,\sigma}(\tau)
   -\mu_{X,\sigma}^{\rm eff} n_{Xi,\sigma}(\tau) ]
\]
\[
   -t \sum_{\langle Ai,Bj \rangle,\sigma} [
  c_{Ai,\sigma}^{\dagger}(\tau) c_{Bj,\sigma}(\tau)+\; {\rm h.c.}]
   +U^{\rm eff} \sum_{Xi} :n_{Xi,\uparrow}(\tau)::n_{Xi,\downarrow}(\tau):
\] 
\begin{equation}
\label{eq-34}
  +\frac{1}{2}\sum_{Xi,Yj} :\phi_{Xi}(\tau): 
  \tilde{V}_{Xi,Yj}^{-1} :\phi_{Yj}(\tau):
  - \sum_{Xi} :\phi_{Xi}(\tau):
  :[n_{Xi,\uparrow}(\tau)+n_{Xi,\downarrow}(\tau)]:
  \},
\end{equation}

\noindent with

\[
  \mu_{X,\sigma}^{\rm eff}=\mu_{X,\sigma}-\frac{1}{2}U
  -\sum_{j} V_{\langle Xi,\overline{X}j \rangle}\langle
  [n_{\overline{X}j,\uparrow}(\tau)+n_{\overline{X}j,\downarrow}(\tau)]
  \rangle,
\]
\[
  U^{\rm eff}=U+\lambda,
\]

\noindent where $\overline{X}=B$ if $X=A$ and vice versa.
The Green's functions we are going to use are defined as follows:

\begin{equation}
\label{eq-35}
  G_{\sigma}^{\rm XY}(i\tau|i^{\prime}\tau^{\prime}) \stackrel{\rm def}{=}
  -\langle {\rm T}_{\tau} c_{{\rm X}i,\sigma}(\tau)
  c_{{\rm Y}i^{\prime},\sigma}^{\dagger}(\tau^{\prime}) \rangle,
\end{equation}
\begin{equation}
\label{eq-36}
  \chi^{\rm XY}(i\tau|i^{\prime}\tau^{\prime}) \stackrel{\rm def}{=}
  -\langle {\rm T}_{\tau}
  :[n_{{\rm X}i,\uparrow}(\tau)+n_{{\rm X}i,\downarrow}(\tau)]:
  :[n_{{\rm Y}i^{\prime},\uparrow}(\tau^{\prime})+n_{{\rm Y}i^{\prime},
  \downarrow}(\tau^{\prime})]: \rangle,
\end{equation}
\begin{equation}
\label{eq-37}
  D^{\rm XY}(i\tau|i^{\prime}\tau^{\prime}) \stackrel{\rm def}{=}
  -\langle {\rm T}_{\tau} :\phi_{{\rm X}i}(\tau):
  :\phi_{{\rm Y}i^{\prime}}(\tau^{\prime}): \rangle.
\end{equation}

\noindent The Dyson equations are now $2 \times 2$ matrix equations:

\begin{equation}
\label{eq-38}
  \left[
    \begin{array}{cc}
      G_{\sigma}^{\rm AA} & G_{\sigma}^{\rm AB} \\
      G_{\sigma}^{\rm BA} & G_{\sigma}^{\rm BB} \\
    \end{array}
  \right]^{-1} (k,ip_n)=
  \left[
    \begin{array}{cc}
      ip_n+\mu_{{\rm A}\sigma} & -t_k  \\
      -t_{-k} & ip_n+\mu_{{\rm B}\sigma}   \\
    \end{array}
  \right]-
  \left[
    \begin{array}{cc}
      \Sigma_{\sigma}^{\rm AA} & \Sigma_{\sigma}^{\rm AB} \\
      \Sigma_{\sigma}^{\rm BA} & \Sigma_{\sigma}^{\rm BB} \\
    \end{array}
  \right] (k,ip_n).
\end{equation}

\begin{equation}
\label{eq-39}
  \left[
    \begin{array}{cc}
      D^{\rm AA} & D^{\rm AB} \\
      D^{\rm BA} & D^{\rm BB} \\
    \end{array}
  \right]^{-1} (k,i\omega_n)
  =-\left[
    \begin{array}{cc}
      \lambda & -2V_k  \\
      -2V_{-k} & \lambda   \\
    \end{array}
  \right]^{-1} -
  \left[
    \begin{array}{cc}
      \Pi^{\rm AA} & \Pi^{\rm AB} \\
      \Pi^{\rm BA} & \Pi^{\rm BB} \\
    \end{array}
  \right](k,i\omega_n)
\end{equation}

\noindent In the above equations and for all those with two sublattices,
we always define the momentum in the reduced Brillouin zone. If the lattice
under consideration is of supercubic type
in d-dimension, one can easily find that for the nearest neighbor
hopping and interaction, $t_k=t\sum_{i=1}^d \cos k_i$ and
$V_k=V\sum_{i=1}^d \cos k_i$. Similar as that for the homogeneous system,
we can derive an identity relating the phonon and the electron density
Green's functions

\begin{equation}
\label{eq-40}
  \left[
    \begin{array}{cc}
      \chi^{\rm AA} & \chi^{\rm AB} \\
      \chi^{\rm BA} & \chi^{\rm BB} \\
    \end{array}
  \right]^{-1} (k,i\omega_n)=
  \left[
    \begin{array}{cc}
      \lambda & -2V_{k}  \\
      -2V_{-k} & \lambda   \\
    \end{array}
  \right]+
  \left[
    \begin{array}{cc}
      \Pi^{\rm AA} & \Pi^{\rm AB} \\
      \Pi^{\rm BA} & \Pi^{\rm BB} \\
    \end{array}
  \right]^{-1} (k,i\omega_n).
\end{equation}

We are now in the position to introduce the E-DMFT approximation.
Following the same procedure as before, we can write down the
effective E-DMFT action by using the cavity construction. What we do
here is that we first integrate out all, including those in both sublattices,
but two neighboring lattice sites, one from each of the two sublattices.
In this way, we keep a cluster containing two representative lattice sites.
The cluster plays the role as a composite impurity which is coupled to the
self-consistent fermionic and bosonic baths. As we have mentioned earlier,
it is found \cite{venky2} that the Hartree terms from the non-local
interaction across the cluster
boundary contribute to the effective action. However, since the 
Hamiltonian we use here is prepared in such a way that there is no longer
Hartree terms contained in the interaction, the procedure towards
E-DMFT becomes very straightforward. The effective action is given by:

\[
  S^{\rm eff}=-\int_0^{\beta} d\tau \int_0^{\beta} d\tau^{\prime}
  \sum_{{\rm XY},\sigma}c_{{\rm X},\sigma}^{\dagger}(\tau)
  [{\cal G}^{\rm XY}_{\sigma}]^{-1}
  (\tau-\tau^{\prime}) c_{{\rm Y},\sigma}(\tau^{\prime})
\]
\[
  -\frac{1}{2}\int_0^{\beta} d\tau \int_0^{\beta} d\tau^{\prime}
  \sum_{\rm XY}:\phi_{\rm X}(\tau):[{\cal D}^{\rm XY}]^{-1}
  (\tau-\tau^{\prime}):\phi_{\rm Y}(\tau^{\prime}):
\]
\begin{equation}
\label{eq-41}
  +U^{\rm eff} \int_0^{\beta} d\tau \sum_{\rm X}
  :n_{{\rm X},\uparrow}(\tau)::n_{{\rm X},\downarrow}(\tau):
  - \int_0^{\beta} d\tau \sum_{\rm X} :\phi_{\rm X}(\tau):
  :[n_{{\rm X},\uparrow}(\tau)+n_{{\rm X},\downarrow}(\tau)]:
\end{equation}

\noindent with ${\rm X,Y}$ summed over ${\rm A,B}$. From the effective
action, we can
measure the impurity Green's functions and calculate the self-energies
by using the local Dyson equations. The self-consistency is reached
by identifying the impurity Green's functions with the local Green's 
functions which are given as follows:

\[
  \left[
    \begin{array}{cc}
      G_{\sigma}^{\rm loc, AA} & G_{\sigma}^{\rm loc, AB} \\
      G_{\sigma}^{\rm loc, BA} & G_{\sigma}^{\rm loc, BB} \\
    \end{array}
  \right](ip_n)= \sum_k
  \left[
    \begin{array}{cc}
      G_{\sigma}^{\rm AA} & G_{\sigma}^{\rm AB} \\
      G_{\sigma}^{\rm BA} & G_{\sigma}^{\rm BB} \\
    \end{array}
  \right](k,ip_n)
\]
\begin{equation}
\label{eq-42}
  =\sum_k \left\{
  \left[
    \begin{array}{cc}
      ip_n+\mu_{{\rm A}\sigma} & t_k  \\
      t_{-k} & ip_n+\mu_{{\rm B}\sigma}   \\
    \end{array}
  \right]-
  \left[
    \begin{array}{cc}
      \Sigma_{\sigma}^{\rm AA}(ip_n) & \Sigma_{\sigma}^{\rm AB}(ip_n) \\
      \Sigma_{\sigma}^{\rm BA}(ip_n) & \Sigma_{\sigma}^{\rm BB}(ip_n) \\
    \end{array}
  \right] \right\}^{-1}
\end{equation}  

\[
  \left[
    \begin{array}{cc}
      D^{\rm loc, AA} & D^{\rm loc, AB} \\
      D^{\rm loc, BA} & D^{\rm loc, BB} \\
    \end{array}
  \right](i\omega_n) = \sum_k
  \left[
    \begin{array}{cc}
      D^{\rm AA} & D^{\rm AB} \\
      D^{\rm BA} & D^{\rm BB} \\
    \end{array}
  \right](k,i\omega_n)
\]
\begin{equation}
\label{eq-43}
  =-\sum_k\left\{
  \left[
    \begin{array}{cc}
      \lambda & -2V_k  \\
      -2V_{-k} & \lambda   \\
    \end{array}
  \right]^{-1} +
  \left[
    \begin{array}{cc}
      \Pi^{\rm AA}(i\omega_n) & \Pi^{\rm AB}(i\omega_n) \\
      \Pi^{\rm BA}(i\omega_n) & \Pi^{\rm BB}(i\omega_n) \\
    \end{array}
  \right] \right\}^{-1}
\end{equation}

\[
  \left[
    \begin{array}{cc}
      \chi^{\rm loc, AA} & \chi^{\rm loc, AB} \\
      \chi^{\rm loc, BA} & \chi^{\rm loc, BB} \\
    \end{array}
  \right](i\omega_n)
  =\sum_k
    \left[
    \begin{array}{cc}
      \chi^{\rm AA} & \chi^{\rm AB} \\
      \chi^{\rm BA} & \chi^{\rm BB} \\
    \end{array}
  \right](k,i\omega_n)
\]
\begin{equation}
\label{eq-44}
  =\sum_k \left\{
  \left[
    \begin{array}{cc}
      \lambda & -2V_k  \\
      -2V_{-k} & \lambda   \\
    \end{array}
  \right]+
  \left[
    \begin{array}{cc}
      \Pi^{\rm AA}(i\omega_n) & \Pi^{\rm AB}(i\omega_n) \\
      \Pi^{\rm BA}(i\omega_n) & \Pi^{\rm BB}(i\omega_n) \\
    \end{array}
  \right]^{-1} \right\}^{-1}.
\end{equation}

\noindent Same as before, the $\lambda$ dependences cancel out exactly.
Combining the last two of the self-consistent equations, (\ref{eq-43}) and
(\ref{eq-44}), we obtain the following identity:

\[
  \left[
    \begin{array}{cc}
      \chi^{\rm loc, AA}(i\omega_n) & \chi^{\rm loc, AB}(i\omega_n) \\
      \chi^{\rm loc, BA}(i\omega_n) & \chi^{\rm loc, BB}(i\omega_n) \\
    \end{array}
  \right]
  =
  \left[
    \begin{array}{cc}
      \Pi^{\rm AA}(i\omega_n) & \Pi^{\rm AB}(i\omega_n) \\
      \Pi^{\rm BA}(i\omega_n) & \Pi^{\rm BB}(i\omega_n) \\
    \end{array}
  \right] \times
\]
\begin{equation}
\label{eq-45}
  \left\{  
  \left[
    \begin{array}{cc}
      1 & 0\\
      0 & 1\\
    \end{array}
  \right]
  +
  \left[
    \begin{array}{cc}
      D^{\rm loc, AA}(i\omega_n) & D^{\rm loc, AB}(i\omega_n) \\
      D^{\rm loc, BA}(i\omega_n) & D^{\rm loc, BB}(i\omega_n) \\
    \end{array}
  \right]
  \left[
    \begin{array}{cc}
      \Pi^{\rm AA}(i\omega_n) & \Pi^{\rm AB}(i\omega_n) \\
      \Pi^{\rm BA}(i\omega_n) & \Pi^{\rm BB}(i\omega_n) \\
    \end{array}
  \right] 
  \right\}
\end{equation}

\noindent It is not difficult to check that the above equation can
be obtained directly from the effective action, Eq.(\ref{eq-41}).
One needs only to write down the phonon Green's
function and then integrate out the auxiliary fields. By using the
local phonon Dyson equation, which is the matrix version of
Eq.(\ref{eq-16}), one recovers Eq.(\ref{eq-45}).\\

This completes the formulation of E-DMFT with two sublattices.
One can see that the theory easily combines E-DMFT with
C-DMFT. We employ here a cluster of two sites, with the application
in mind which will be discussed in \S\ref{sec-07}. There is, however,
no difficulty to extend the formalism to clusters of any size.\\

\section{GW Method Combined with E-DMFT}
\label{sec-06}

As we have mentioned, C-DMFT \cite{kotliar2001} allows to pick out
a representative lattice cluster, instead of a single site, in order
to describe a many body system. This makes possible to
treat the finite range interaction as well as the broken symmetry
phase {\it within} the cluster. The advantage of C-DMFT is that
it solves exactly the cluster so that the spatially non-local
correlations within the cluster are automatically taken into
account. In combining with E-DMFT, C-DMFT is also able to handle
interactions with range beyond the cluster size, as we have shown
in the last section. However, the price one has to pay is that in
solving a cluster, a lot more technical resources are needed.\\

\begin{figure}[h]
  \epsfxsize=14.0cm
  \centerline{\epsfbox{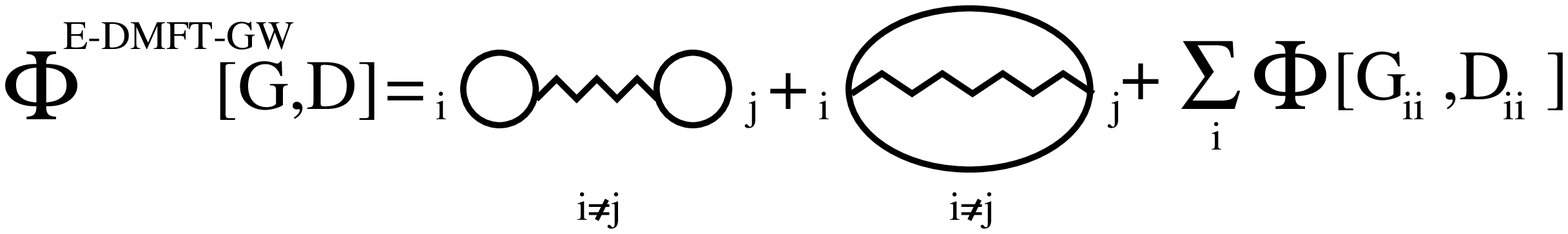}}
  \caption{The 2-particle irreducible functional $\Phi[G,D]$ in the E-DMFT-GW
  approach.}
\label{fig-17}
\end{figure}

In this section we propose a less computationally intensive 
prescription as compared to the E-DMFT + C-DMFT procedure. It is
based on the following physical idea. In real materials, the
on-site Hubbard interaction $U$ is much larger than the non-local
ones. Hence the local interaction has to be treated
non-perturbatively (namely with DMFT) in order to obtain the
local self-energy. Meanwhile it is legitimate to
make perturbative expansion to obtain the non-local part of the
self-energy in the spirit of the GW method \cite{hedin}. The
original GW method computes a screened Coulomb line W by summing
Random Phase Approximation (RPA) diagrams and obtains the one
electron self-energy by
considering the lowest order graph in W, hence the name GW.
The E-DMFT-GW approach is derivable via the Baym-Kadanoff
functional \cite{gabi}. The functional derivatives of the
2-particle irreducible part of the Baym-Kadanoff functional,
$\Phi(G,D)$, with respect to the full Green's function give the
corresponding self-energies. $\Phi(G,D)$ is constructed with
the full Green's functions $G$ and $D$ and the interaction
vertices (As discussed in \S\ref{sec-08}, the choice of the
phonon field should be done judiciously). The
E-DMFT-GW method consists of approximating $\Phi$ (see
Fig.\ref{fig-17}) by the leading order non-local graphs
and evaluating the rest of the functional
$\Phi$ in the local approximation.
The E-DMFT-GW self-energies are given by

\[
   \Sigma_{\sigma;i,j}(G,D)=\frac{\delta \Phi(G,D)}{\delta G_{\sigma;i,j}}
   \simeq \delta_{ij}\Sigma_{\sigma;i,i}^{\rm E-DMFT}(G,D)
   +(1-\delta_{ij})\Sigma_{\sigma;i,j}^{\rm GW}(G,D)
\]
\begin{equation}
\label{eq-add-1}
  \Pi_{i,j}(G,D)=\frac{\delta \Phi(G,D)}{\delta D_{i,j}}
  \simeq \delta_{ij}\Pi_{i,i}^{\rm E-DMFT}(G,D)
  +(1-\delta_{ij})\Pi_{i,j}^{\rm GW}(G,D)
\end{equation}

\noindent For the approach to be derivable from a functional, $\Sigma(G,D)$
and $\Pi(G,D)$ has to be calculated self-consistently. The GW method has
been applied in {\it ab initio} calculations of semiconductors since
the original works by G. Strinati, H.J. Mattausch, and W. Hanke \cite{hanke}.
However, it was pointed out that in LDA-GW for electron systems \cite{holm}
the self-consistency results in incorrect one-electron spectra and that
it is better to compute $\Sigma(G_0)$ with $G_0$ the unperturbed Green's
function instead. Better total energy, though, is obtained from $\Sigma(G)$
\cite{holm,leeuwen}. We believe that our proposal resolves this
contradiction. The E-DMFT iteration obtains the largest self-energy term
(the on-site one) self-consistently and non-perturbatively. The GW
approximation is used for the smaller term (the off-diagonal one). In
our model calculation we find the difference between E-DMFT +
non-self-consistent GW and E-DMFT + self-consistent GW is small.
This can be generalized in a straightforward way to realistic multiband
situations. \\

\begin{figure}[h]
  \epsfxsize=14.0cm
  \centerline{\epsfbox{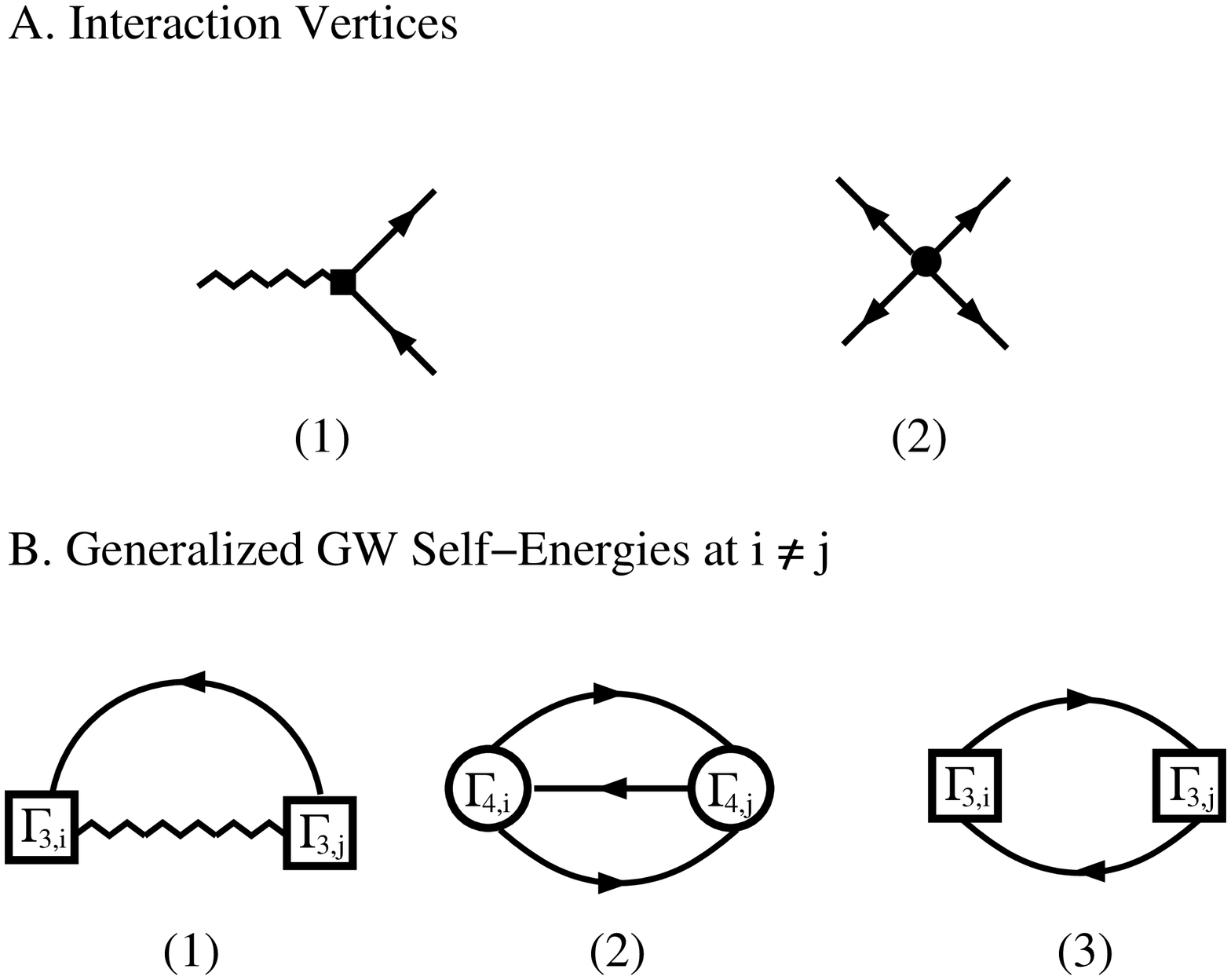}} 
  \caption{ A. The bare interaction vertices contained in
    Eqs.(\ref{eq-06}) and (\ref{eq-34}), that is, the local electron-phonon
    interaction A(1) and the local Hubbard interaction A(2). 
   B. The non-local self-energy contributions described in
   Eq.(\ref{eq-48}) [B(1) and B(2)] for electrons and
   Eq.(\ref{eq-49}) [B(3)] for phonons. The strengths of the leading
   contributions (from the nearest neighbors) for the three diagrams are 
   $O[V/d]$, $O[U^2/d^{3/2}]$, and $O[1/d]$, respectively. Since in
   each of the diagrams we require the vertices be from different lattice
   sites, there is no double counting.
   }
\label{fig-18}
\end{figure}

We need to discuss more specifically the non-local self-energy
diagrams in our generalized GW approach in combination with the
E-DMFT. We identify two such contributions.
The first is the boson exchange diagram which is of the
same form as that in the GW method  [Fig.\ref{fig-18}(B1)]. We
require that the two vertices $\Gamma_3$ be local and come
from different lattice sites, giving rise to the off-diagonal
self-energy. Fig.\ref{fig-18}(B1) uses the full electron-phonon
vertices, instead of the bare ones, in the exchange diagram of
the self-energy. Those local vertices, which can be measured in
E-DMFT-QMC, are defined through the following
Green's function \cite{abrikosov}:

\[
  G^{\rm loc}
  (\tau_1\sigma_1,\tau_2\sigma_2;\tau_3)
  \stackrel{\rm def}{=}
  \langle {\rm T}_{\tau} c_{\sigma_1}(\tau_1)
  c_{\sigma_2}^{\dagger}(\tau_2)\phi(\tau_3)\rangle
\]
\[
   =\int_0^{\beta} d\tau_1^{\prime} \sum_{\sigma_1^{\prime}}
   \int_0^{\beta} d\tau_2^{\prime} \sum_{\sigma_2^{\prime}}
   \int_0^{\beta} d\tau_3^{\prime}
   G^{\rm loc}(\tau_1\sigma_1|\tau_1^{\prime}\sigma_1^{\prime})
   G^{\rm loc}(\tau_2\sigma_2|\tau_2^{\prime}\sigma_2^{\prime})
\]
\begin{equation}
\label{eq-46}
  \Gamma_3^{\rm loc}(\tau_1^{\prime}\sigma_1^{\prime},
  \tau_2^{\prime}\sigma_2^{\prime};\tau_3^{\prime})
  D^{\rm loc}(\tau_3^{\prime}|\tau_3).
\end{equation}

\noindent Unlike the skeleton diagram commonly used for the
electron self-energy \cite{abrikosov} where one of the vertices
should be bare to avoid over-counting, the diagram
Fig.\ref{fig-18}(B1) uses two full local vertices. Our
requirement, that the two vertices be from different lattice
sites, ensures that there is no over-counting. While the skeleton
construction uses one bare and one full vertex to produce the
exact self-energy, our method uses two full local vertices
to produce the leading non-local correction. We should also remark
that both the electron and phonon lines appeared in
Fig.\ref{fig-18}(B1) represent the non-perturbative Green's
functions from E-DMFT. Especially, the phonon Green's
function plays the role as the screened Coulomb interaction
which in the original GW is obtained by using RPA.\\

There is another contribution, which originates from the local
interaction $U$ at the second order [Fig.\ref{fig-18}(B2)]. The
reason this contribution is important is that usually $U$ is
much bigger than $V$. The effective local vertex $\Gamma_4^{\rm loc}$
is defined in the following way:

\[
  G^{(2),{\rm loc}}
  (\tau_1\sigma_1,\tau_2\sigma_2|\tau_3\sigma_3,\tau_4\sigma_4)
  \stackrel{\rm def}{=}
  \langle {\rm T}_{\tau} c_{\sigma_1}(\tau_1) c_{\sigma_2}(\tau_2)
  c_{\sigma_3}^{\dagger}(\tau_3) c_{\sigma_4}^{\dagger}(\tau_4) \rangle
\]
\[
  =G^{\rm loc}(\tau_1\sigma_1|\tau_4\sigma_4)
   G^{\rm loc}(\tau_2\sigma_2|\tau_3\sigma_3)
   -G^{\rm loc}(\tau_2\sigma_2|\tau_4\sigma_4)
   G^{\rm loc}(\tau_1\sigma_1|\tau_3\sigma_3)
\]
\[
  +\int_0^{\beta} d\tau_1^{\prime} \sum_{\sigma_1^{\prime}}
   \int_0^{\beta} d\tau_2^{\prime} \sum_{\sigma_2^{\prime}}
   \int_0^{\beta} d\tau_3^{\prime} \sum_{\sigma_3^{\prime}}
   \int_0^{\beta} d\tau_4^{\prime} \sum_{\sigma_4^{\prime}}
   G^{\rm loc}(\tau_1\sigma_1|\tau_1^{\prime}\sigma_1^{\prime})
   G^{\rm loc}(\tau_2\sigma_2|\tau_2^{\prime}\sigma_2^{\prime})
\]
\begin{equation}
\label{eq-47}
  \Gamma_4^{\rm loc}(\tau_1^{\prime}\sigma_1^{\prime},
  \tau_2^{\prime}\sigma_2^{\prime}|\tau_3^{\prime}\sigma_3^{\prime},
  \tau_4^{\prime}\sigma_4^{\prime})
  G^{\rm loc}(\tau_3^{\prime}\sigma_3^{\prime}|\tau_3\sigma_3)
  G^{\rm loc}(\tau_4^{\prime}\sigma_4^{\prime}|\tau_4\sigma_4)
\end{equation}

\noindent If there is no external magnetic field, only two
spin configurations are allowed in the two particle Green's function
as well as the vertex, that with all the spins in the same direction
and that with two spins up and two spins down \cite{nozieres}.
Once we get the local vertex, the corresponding contribution to the
off-site self-energy can be constructed as shown in Fig.\ref{fig-18}(B2),
with again the vertices coming from different sites and ensuring
no double counting of the diagrams. It should also be pointed out that
the diagrams contained in Fig.\ref{fig-18}(B1) and (B2) are totally
different. This can be seen easily by comparing the non-local lines
in the two diagrams.\\

To compare the importance of the two terms, we can investigate
their scaling behavior with the spatial dimension \cite{metzner}. We
should keep in mind that both G and W scale as $1/\sqrt{d}$ for
nearest neighbors in real space,
with $d$ the dimension, and both the interaction vertices are
local which means they do not scale. We then see that the leading
electron-phonon contribution scales as $1/d$ and that of
the on-site interaction as $1/d^{3/2}$. In the infinite
dimension limit there is no doubt that the electron-phonon
contribution is more important. However, if we work in
finite spatial dimensions (usually $\le 3$) and since the
on-site interaction $U$ is likely much bigger than the off-site
one, $V$, these two can be of the same order practically. This
actually happens in the example in 1-D which we will show in the
next section.\\

In the same spirit one can also obtain the leading non-local
correction to the phonon self-energy, {\it i.e.} Fig.\ref{fig-18}B(3),
which is the leading non-local correction in
terms of the E-DMFT interactions. Due to the two electron lines,
the diagram scales as $1/d$ when the two contributing lattice
sites are nearest neighbors.\\

In practical calculations, we first solve E-DMFT iteratively and
obtain all the local self-energies and the interaction vertices.
We then apply GW approximation
to calculate the non-local self-energies by using the Green's
functions obtained from E-DMFT. As in the original GW, we assume
here that the corrections do not change dramatically the physical
properties of the system so that we are allowed to use the quantities
from E-DMFT directly. The last step is to use the approximate
self-energies in the exact Dyson equation so that
all the Green's functions, with both spatial and temporal
dependences, can be calculated.\\

The GW contributions to the self-energies are
calculated in the real space-time as follows for $i \neq j$
(see Fig.\ref{fig-18})

\[
  \Sigma_{\sigma}^{\rm GW}(i\tau|j\tau^{\prime})
  =-\int_0^{\beta} d\tau_1 \sum_{\sigma_1}
  \int_0^{\beta} d\tau_1^{\prime}\sum_{\sigma_1^{\prime}}
  \int_0^{\beta} d\tau_2 \int_0^{\beta} d\tau_2^{\prime}
\]
\[
  \Gamma_{3,i}^{\rm loc}(\tau\sigma,\tau_1\sigma_1;\tau_2)
  G(i\tau_1\sigma_1|j\tau_1^{\prime}\sigma_1^{\prime})
  D(i\tau_2|j\tau_2^{\prime})
  \Gamma_{3,j}^{\rm loc}(\tau_1^{\prime}\sigma_1^{\prime},
  \tau^{\prime}\sigma;\tau_2^{\prime})
\]
\[
  -\int_0^{\beta} d\tau_1 \sum_{\sigma_1}
   \int_0^{\beta} d\tau_1^{\prime} \sum_{\sigma_1^{\prime}}
   \int_0^{\beta} d\tau_2 \sum_{\sigma_2}
   \int_0^{\beta} d\tau_2^{\prime} \sum_{\sigma_2^{\prime}}
   \int_0^{\beta} d\tau_3 \sum_{\sigma_3}
   \int_0^{\beta} d\tau_3^{\prime} \sum_{\sigma_3^{\prime}}
\]
\begin{equation}
\label{eq-48}
  \Gamma_{4,i}^{\rm loc}(\tau\sigma,\tau_1\sigma_1|
  \tau_2\sigma_2,\tau_3\sigma_3)
  G(i\tau_1\sigma_1|j\tau_1^{\prime}\sigma_1^{\prime})
  G(i\tau_2\sigma_2|j\tau_2^{\prime}\sigma_2^{\prime})
  G(j\tau_3^{\prime}\sigma_3^{\prime}|i\tau_3\sigma_3)
  \Gamma_{4,j}^{\rm loc}(\tau_3^{\prime}\sigma_3^{\prime},
  \tau_2^{\prime}\sigma_2^{\prime}|\tau_1^{\prime}\sigma_1^{\prime},
  \tau^{\prime}\sigma)
\end{equation}

\[
  \Pi^{\rm GW}(i\tau|j\tau^{\prime})
  =\int_0^{\beta} d\tau_1 \sum_{\sigma_1}
  \int_0^{\beta} d\tau_1^{\prime}\sum_{\sigma_1^{\prime}}
  \int_0^{\beta} d\tau_2 \sum_{\sigma_2}
   \int_0^{\beta} d\tau_2^{\prime} \sum_{\sigma_2^{\prime}}
\]
\begin{equation}
\label{eq-49}
   \Gamma_{3,i}^{\rm loc}(\tau_1\sigma_1,\tau_2\sigma_2;\tau)
   G(i\tau_1\sigma_1|j\tau_1^{\prime}\sigma_1^{\prime})
   G(j\tau_2^{\prime}\sigma_2^{\prime}|i\tau_2\sigma_2)
   \Gamma_{3,j}^{\rm loc}(\tau_2^{\prime}\sigma_2^{\prime},
   \tau_1^{\prime}\sigma_1^{\prime};\tau^{\prime})
\end{equation}

\noindent In the above equations we labeled the vertices by the
lattice site index with in mind the possibility of inequivalent
sublattices. There is a symmetry one can use in the calculation
at $i \neq j$:

\begin{equation}
\label{eq-50}
  G_{ \sigma}(i\tau|j\tau^{\prime})=
  [G_{ \sigma}(j\tau^{\prime}|i\tau)]^{\dagger}
\end{equation}

\noindent It is both physically transparent and technically convenient
to perform the generalized GW calculation in the coordinate space and
imaginary time \cite{godby}. It is also very easy to extend the
expressions to the systems with different sublattices, as we will show
in the next section.\\

\section{Application II: 1D Band Insulator via E-DMFT plus GW}
\label{sec-07}

Since we want to investigate if the GW method can improve the E-DMFT
results, we need to know the corresponding exact solution of the model
under investigation. In this section, we implement the E-DMFT of two
sublattices for a 1D U-V model with an alternating chemical potential. 
This model can be solved exactly at $T=0$ via Density Matrix
Renormalization Group (DMRG) \cite{venky0,venky1}.
The model is relevant in the study of the interplay between the
electronic correlation and the electron-phonon coupling in the
mixed stack organic compounds \cite{torrance,nagaosa} and the
ferroelectric perovskites \cite{egami}. The phase diagram has been
studied \cite{torio}.\\

Because we solve the impurity model in E-DMFT using QMC simulation
which works at finite temperatures, we need to make the comparison in
the band insulating phase where all the excitations are gapped so that
the effects of the thermal excitations are suppressed at low enough
temperatures. We also want to make the comparison in the parameter space
where the quantum fluctuation is strong enough so that the standard
mean field solution does not work. It is known that in such a case one
needs C-DMFT of at least two sites to get good agreement with the exact
solution \cite{venky2}. Our goal here, though, is mainly to see if the
GW method can improve the E-DMFT results.\\

Due to the above reasons, the E-DMFT we are going to use in this section
is slightly different from that described in \S\ref{sec-05} which
combined E-DMFT with C-DMFT. The {\it pure} E-DMFT (without
C-DMFT) for two sublattices is established in the following way. We
first choose a representative site from sublattice A, integrate out all
the other sites in both the sublattices, and obtain an effective impurity
action for this site. Then, from the nearest neighbors of this site, we
choose another representative site, which obviously belongs to
sublattice B, and repeat the same procedure. The two effective actions
reached in this way are the same as that for a homogeneous system we
described in \S\ref{sec-03}. They are formally independent to each
other at the level of the impurity model. Of course the two are
connected, at the self-consistency, through the Dyson equations which
are the same as those given by Eqs. (\ref{eq-38}) and (\ref{eq-39}),
except the
off-diagonal self-energies are now zero. Technically, one can easily
understand the structure of this E-DMFT by imposing the requirement on
all the corresponding equations in \S\ref{sec-05} that the impurity
model be restrictively local. Then all the off-diagonal dynamical Weiss
fields, and thus the off-diagonal self-energies, are zero. However, the
impurity Green's functions still have non-local contributions, as is
evident from Eqs.(\ref{eq-42})-(\ref{eq-44}). This scenario of
implementing (E-) DMFT, in midway between the single site (E-) DMFT and
the cluster one, has been used successfully in treating systems with
inequivalent sublattices while avoiding the heavy calculations needed
in C-DMFT \cite{kotliar1996}.\\

We study here the following Hamiltonian:

\[
  \hat{H}= -t \sum_{i,\sigma} (
  \hat{c}_{i,\sigma}^{\dagger}\hat{c}_{i+1,\sigma}+\; {\rm h.c.})
  -\sum_{i,\sigma} \mu_i \hat{n}_{i,\sigma}
\]
\begin{equation}
\label{eq-51}
  +U\sum_i (\hat{n}_{i,\uparrow}-\frac{1}{2})
           (\hat{n}_{i,\downarrow}-\frac{1}{2})
  +V\sum_{i} 
  (\hat{n}_{i,\uparrow}+\hat{n}_{i,\downarrow}-1)
  (\hat{n}_{i+1,\uparrow}+\hat{n}_{i+1,\downarrow}-1).
\end{equation}

\noindent We consider the special case with alternating chemical potential:

\begin{equation}
\label{eq-52}
   \mu_i= (-1)^i \mu
\end{equation}

\noindent In this case we know the exact forms of the hopping
matrix element and the non-local interaction. We can write down
for the off-diagonal terms in the Dyson equations (\ref{eq-38})
and (\ref{eq-39})

\begin{equation}
\label{eq-53}
  t_k=te^{ik}\cos k,
  \;\;\;\;\;
  V_k=Ve^{ik}\cos k
\end{equation}

\noindent  The momentum $k$ is again restricted in the reduced
Brillouin zone, given by  $-\pi /2<k \le \pi/2$ in the current
case. The phase factors $\exp(\pm ik)$ come from the
fact that each site (A,i) [(B,i)] has two neighbors, one within the same
unit cell, (B,i) [(A,i)] and the other comes from the cell to the left
(right), (B,i-1) [(A,i+1)]. Hence the later contributes a momentum
dependent phase factor. Remember when we use the above two
equations for self-consistency in E-DMFT, the off-diagonal self-energies
for both electrons and phonons should be set to zero due the assumption of
locality:

\begin{equation}
\label{eq-54}
  \Sigma_{\sigma}^{\rm AB}(ip_n)=0,
  \;\;\;
  \Pi^{\rm AB}(i\omega_n)=0.
\end{equation}

\noindent Because of the dimensionality, the momentum
summations needed in calculating the local Green's functions
can be carried out exactly, which give:

\begin{equation}
\label{eq-55}
  G_{\sigma}^{\rm loc,XX}(ip_n)= \sum_k G_{\sigma}^{\rm XX}(k,ip_n)
  =\frac{\zeta_{\sigma}^{\rm \overline{XX}}(ip_n)}{
  \zeta_{\sigma}^{\rm AA}(ip_n)
  \zeta_{\sigma}^{\rm BB}(ip_n)-1}
\end{equation}

\noindent with

\[
  \zeta_{\sigma}^{\rm X}(ip_n)
  \stackrel{\rm def}{=}
  ip_n+\mu_{\rm X}^{\rm eff}-\Sigma^{\rm XX}_{\sigma}(ip_n).
\]

\noindent and

\[
  D^{\rm loc,XX}(i\omega_n) =
  \sum_k D^{\rm XX}(k,i\omega_n)
\]
\begin{equation}
\label{eq-56}
  = \frac{4V^2\Pi^{\rm \overline{XX}}}{
  \sqrt{1-4V^2\Pi^{\rm AA}(i\omega_n) \Pi^{\rm BB}(i\omega_n)}
  [1+\sqrt{1-4V^2\Pi^{\rm AA}(i\omega_n) \Pi^{\rm BB}(i\omega_n)}]}
\end{equation}

\noindent 
In the above, we used again the notation that $\overline{\rm X}=$B
if X$=$A
and vice versa. As we have noted, the solution of the impurity model
in the current case consists of two independent parts, one for
each representative lattice site, and each of them are exactly the
same as that for a homogeneous system. The only difference comes in
at the self-consistent conditions which are given by the above pair
of equations.\\

After we get the solution of the impurity model, either within a
single iteration or after the convergence of the E-DMFT iterations,
we can perform the GW perturbative calculations. To illustrate the
idea and see qualitatively how the GW self-energy can improve the
results, we consider here the simplest and the most important GW
contributions, those from the nearest neighbors. They contribute
directly to the off-diagonal self-energies, while for the diagonal
ones, we use those from the E-DMFT calculation. All the other
contributions are neglected because the Green's functions decay
exponentially as the spatial separation increases in the band
insulating phase.\\

We can now write down the expression for the self-energy matrices as
follows:

\begin{equation}
\label{eq-57}
  \left[
    \begin{array}{cc}
      \Sigma_{\sigma}^{\rm AA} & \Sigma_{\sigma}^{\rm AB} \\
      \Sigma_{\sigma}^{\rm BA} & \Sigma_{\sigma}^{\rm BB} \\
    \end{array}
  \right] (k,ip_n)
  \simeq
  \left[
    \begin{array}{cc}
      \Sigma_{\sigma}^{\rm DMFT,AA}(ip_n) 
      & \Sigma_{\sigma}^{\rm GW,AB}(ip_n)e^{ik}\cos k \\
      \Sigma_{\sigma}^{\rm GW,BA}(ip_n)e^{-ik}\cos k 
      & \Sigma_{\sigma}^{\rm DMFT,BB}(ip_n) \\
    \end{array}
  \right]
\end{equation}

\begin{equation}
\label{eq-58}
  \left[
    \begin{array}{cc}
      \Pi^{\rm AA} & \Pi^{\rm AB} \\
      \Pi^{\rm BA} & \Pi^{\rm BB} \\
    \end{array}
  \right] (k,i\omega_n)
   \simeq
   \left[
    \begin{array}{cc}
      \Pi^{\rm DMFT,AA}(i\omega_n) 
      & \Pi^{\rm GW,AB}(i\omega_n)e^{ik}\cos k \\
      \Pi^{\rm GW,BA}(i\omega_n)e^{-ik}\cos k
      & \Pi^{\rm DMFT,BB}(i\omega_n) \\
    \end{array}
  \right]
\end{equation}

\noindent The momentum dependences of the off-diagonal terms in the
above two equations come in for the same reason as those in the free
electron and phonon propagators. In the current situation both
the GW terms we discussed in the previous section contribute as given
by Eqs.(\ref{eq-48}) and (\ref{eq-49}). To make life easier, we make a
further approximation which replaces the full interaction vertices
by their bare values:

\begin{equation}
\label{eq-59}
  \Gamma^{\rm loc}_3(\tau_1\sigma_1,\tau_2\sigma_2;\tau_3)
  =\delta_{\sigma_1,\sigma_2} \delta(\tau_1-\tau_2) \delta(\tau_1-\tau_3),
\end{equation}
\begin{equation}
\label{eq-60}
  \Gamma_4^{\rm loc}(\tau_1\sigma_1,\tau_2\sigma_2|
  \tau_3\sigma_3,\tau_4\sigma_4)
  =U\delta_{\sigma_1,\sigma_4}\delta_{\sigma_2,\sigma_3}
  \delta(\tau_1-\tau_2) \delta(\tau_1-\tau_3) \delta(\tau_1-\tau_4).
\end{equation}

\noindent Then Eqs.(\ref{eq-48}) and (\ref{eq-49}) are
greatly simplified and give:

\begin{equation}
\label{eq-61}
  \Sigma_{\sigma}^{\rm GW}(i\tau|j\tau^{\prime})
  =-G_{\sigma}(i\tau|j\tau^{\prime}) D(i\tau|j\tau^{\prime})
  -U^2  G_{-\sigma}(i\tau|j\tau^{\prime})
  G_{-\sigma}(j\tau^{\prime}|i\tau)G_{\sigma}(i\tau|j\tau^{\prime}),
\end{equation}
\begin{equation}
\label{eq-62}
  \Pi^{\rm GW}=\sum_{\sigma}
  G_{\sigma}(i\tau|j\tau^{\prime})G_{\sigma}(j\tau^{\prime}|i\tau).
\end{equation}

\noindent In reaching the numerical results we are going to present, we use
$\lambda=2.0$ for the positive-definiteness of the effective interaction
matrix. In every iteration, the impurity model is solved via QMC by
$10^6$ sweeps. To reach the E-DMFT convergence, ten iterations are
usually needed.\\

\begin{figure}[h]
  \epsfxsize=14.0cm
  \centerline{\epsfbox{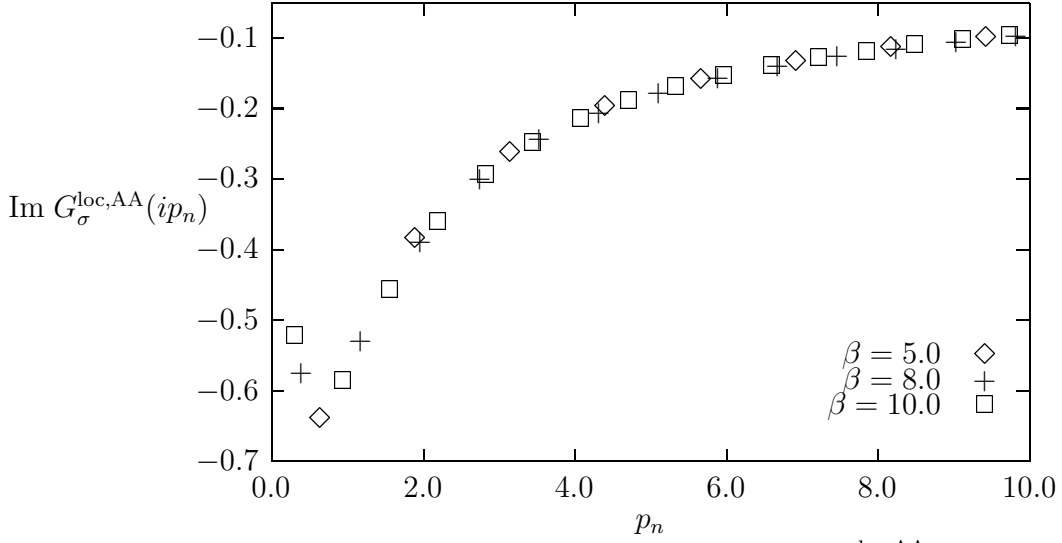}}
  \caption{The imaginary part of the local electron Green's function
     $G_{\sigma}^{\rm loc,AA}(ip_n)$ at $U=5.0$, $V=0.5$, and
     $\mu=2.0$. The results at three different inverse
     temperatures are shown \cite{note3}. Within the
     accuracy of the calculation, the three sets of data
     lie on a smooth curve which means the thermal effects on
     the result has already been suppressed due to the band gap.}
\label{fig-19}
\end{figure}

We first show in Fig.\ref{fig-19} the temperature dependence of the
imaginary part of the electron Green's function calculated at $U=5.0$,
$V=0.5$, and $\mu=2.0$. The three cases are calculated at three different
inverse temperatures, $\beta=5.0, 8.0, 10.0$, with the corresponding
$\Delta\tau=0.25, 0.20, 0.25$, respectively. It is obvious that the three
sets of the data lie on a single smooth curve. Actually this same
situation happens for all the other quantities we measured which are not
shown here. All these suggest that, at the given temperatures, with the
highest at $T=1/5.0$, the thermal fluctuations are already suppressed due
the band gap and we need not worry about the temperature effects. In
the following, we present the results calculated at $\beta=5.0$.\\

\begin{figure}[h]
  \epsfxsize=14.0cm
  \centerline{\epsfbox{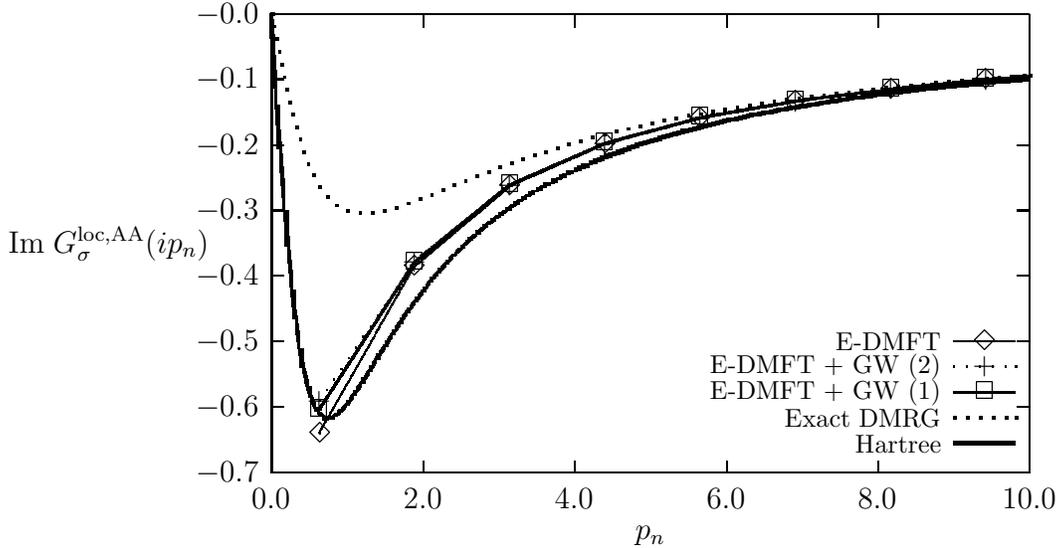}}
  \caption{ 
     The imaginary part of the local electron Green's function
     $G_{\sigma}^{\rm loc,AA}(ip_n)$ at $U=5.0$, $V=0.5$, $\mu=2.0$,
     and $\beta=5.0$. The data labeled as GW(1) comes from the
     contribution described by Fig.\ref{fig-18}B(1) and GW(2)
     from Fig.\ref{fig-18}B(2). It can be seen that GW(1) and
     GW(2) makes corrections at the same order to the E-DMFT
     result.}
\label{fig-20}
\end{figure}

In Fig.\ref{fig-20} we present the data of the imaginary part
of the electron Green's functions calculated using E-DMFT alone,
using GW with the electron-phonon vertex [corresponding to
Fig.\ref{fig-18}B(1)], and with the local Hubbard vertex
[corresponding to Fig.\ref{fig-18}B(2)], respectively. In the
latter two the GW calculations are performed after E-DMFT convergence.
The exact result and the Hartree mean field (MF) result are
also plotted as references. From the results, one can see that
the two terms in the GW correction are of the same order.\\

\begin{figure}[h]
  \epsfxsize=14.0cm
  \centerline{\epsfbox{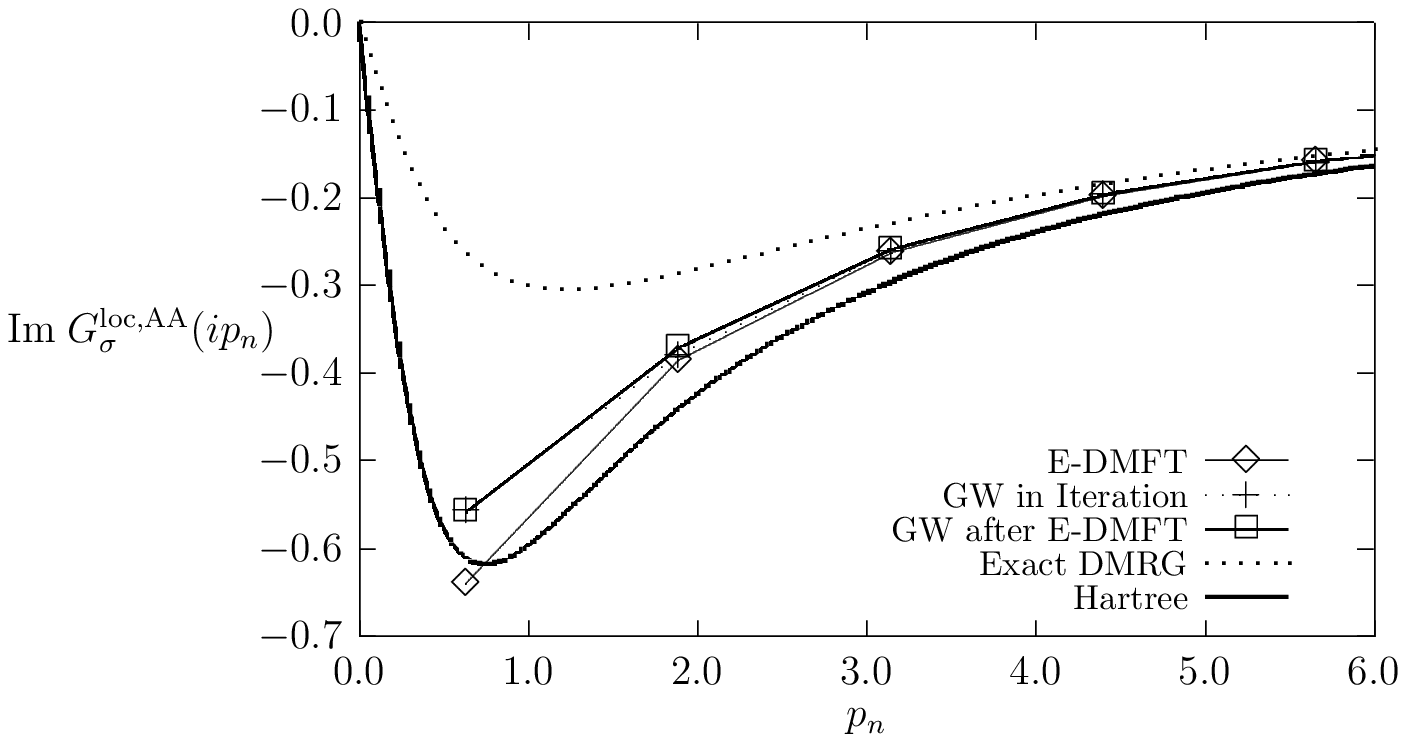}}
  \caption{ 
     The imaginary part of the local electron Green's function
     $G_{\sigma}^{\rm loc,AA}(ip_n)$ at $U=5.0$, $V=0.5$, $\mu=2.0$,
     and $\beta=5.0$.}
\label{fig-21}
\end{figure}

\begin{figure}[h]
  \epsfxsize=14.0cm
  \centerline{\epsfbox{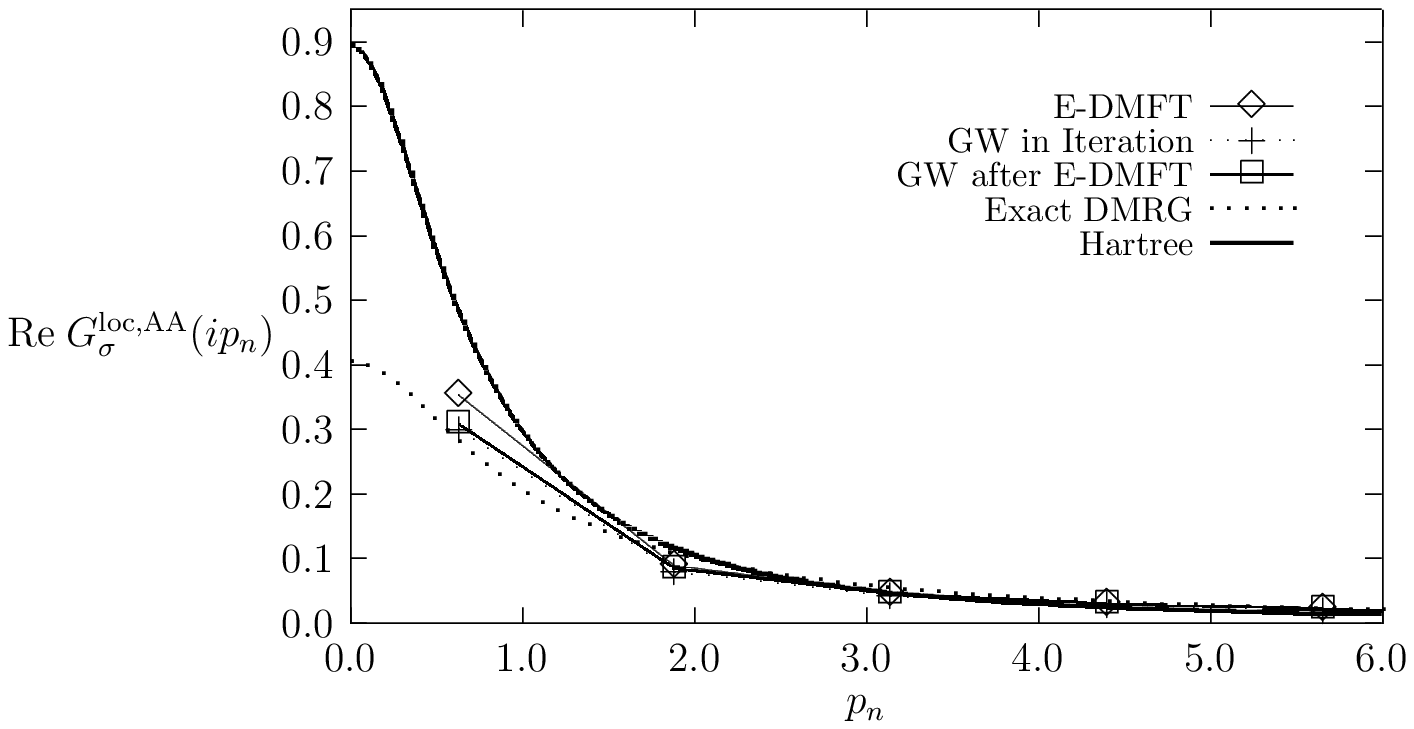}}
  \caption{ 
     The real part of the local electron Green's function
     $G_{\sigma}^{\rm loc,AA}(ip_n)$ at $U=5.0$, $V=0.5$, $\mu=2.0$,
     and $\beta=5.0$.}
\label{fig-22}
\end{figure}

\begin{figure}[h]
  \epsfxsize=14.0cm
  \centerline{\epsfbox{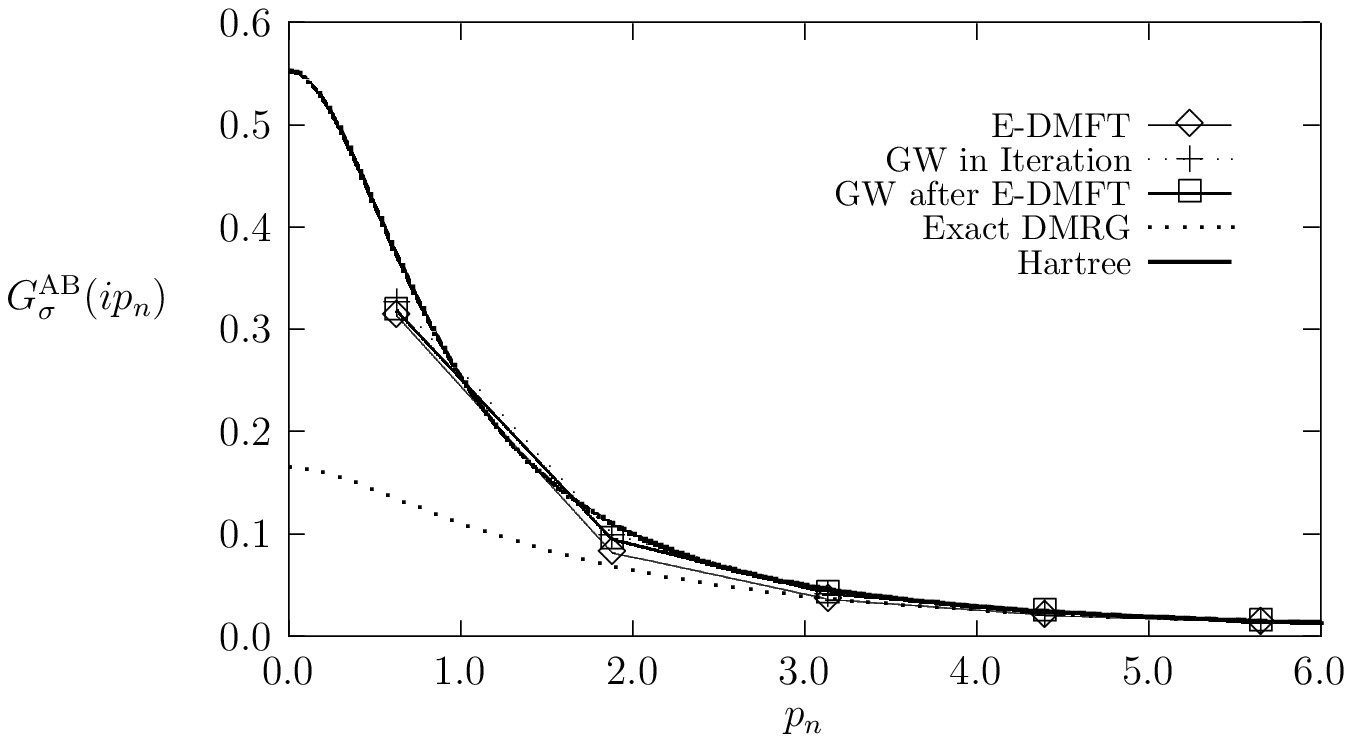}}
  \caption{ 
     The off-diagonal electron Green's function
     $G_{\sigma}^{\rm loc,AB}(ip_n)$ at $U=5.0$, $V=0.5$, $\mu=2.0$,
     and $\beta=5.0$.}
\label{fig-23}
\end{figure}

In Fig.\ref{fig-21} we show the results of the imaginary part
of the local electron Green's function from GW calculation after
E-DMFT convergence and those using GW within the E-DMFT iteration loop.
One can see that the difference is very small. The corresponding real
part is plotted in Fig.\ref{fig-22}. In Fig.\ref{fig-23} we
show the plotting of the Green's function between a pair of neighboring
sites. We can also compare the
result on the average energy per site $\epsilon$. The result of GW after
E-DMFT gives $\epsilon=-1.76$ and that for GW within E-DMFT
$\epsilon=-1.74$. The DMRG finds the exact average ground state energy
per site to be $\epsilon=-2.09$. The difference between the two 
GW+DMFT procedures
is again very minor. From the given results and those performed at the
other parameter points which are not shown here, we conclude that the
two procedures, one with GW after E-DMFT and the other with GW in E-DMFT
iteration, give very close results, although it seems the former is a
little better.\\

The physical information contained in Figs.\ref{fig-20}-\ref{fig-23}
can be understood as follows. (i) In one dimension, all the allowed
modes of the low energy excitations are those bosonic particle hole
pairs carrying momentum $k \sim 0$ and $k \sim 2k_F$ ($=\pi$ at
half-filling) with respect to the Hartree ground state. This explains
why a C-DMFT calculation with a cluster of only two sites gives quite
good results \cite{venky2} while the E-DMFT we employed here does not
work very well at low frequencies.
The difference is basically that a model of a single site can only
capture those modes $k \sim 0$; but a cluster of two sites is already
good enough for those at $k \sim \pi$. (ii) Since the classical Hartree
energy gap,
which is given by $-U/2+2V+\mu=0.5$ is quite small in this case
and both the interactions in Eq.(\ref{eq-51}) are (marginally) relevant
with respect to the metallic Gaussian fixed point,
as we go to lower energy scales and thus longer wave lengths, the
energy gap gets renormalized significantly. This explains why the
exact DMRG result is so different from the MF result at low frequencies.
The high frequency behavior, on the other hand, can be captured fairly well
even by the Hartree approximation, which is evident from
Figs.\ref{fig-20}-\ref{fig-23}. This is the region where all the different
approaches converge to give the same result. (iii) From the figures of
the local electron Green's function one can see that
at not too low frequencies (basically, those beyond the first two Matsubara
frequencies), the E-DMFT result is much better than the MF result and
closer to the exact ones. This
is consistent with the scaling picture since the contributions
to the results at those frequencies higher than the gap energy, which
is of the order of ``$1$" in the current case, can only be from the local
behavior and are described fairly well by the E-DMFT. On the other hand,
this same reason explains the big deviation of the E-DMFT result
at the first two Matsubara frequencies: They are affected more strongly
by those quantum/thermal fluctuations with longer wave length, which are
mostly neglected by the E-DMFT approximation. (iv) The GW works in the way
we have anticipated. It incorporates more spatial correlations into
the self-energy so that the low frequency behavior benefits a lot
from the
correction. As we can see from Figs.\ref{fig-21} and \ref{fig-22}
the GW contributes to both the real and imaginary parts of the local
Green's functions corrections of more than $15\%$. (v) The GW method 
used here has little effect on the nearest neighbor Green's functions
because it is designed to improve the local Green's functions.\\

What we have shown above is that the GW method can be used to
improve the E-DMFT results. Due to the dimensionality, the
single-site E-DMFT results deviates from the exact ones at the
lowest frequencies. By incorporating the spatial correlations,
the GW perturbation contributes a desired correction. Of course
in this very case in 1D it is known that the spatial correlation is
so important that a leading order perturbation is not enough to
recover the exact results. What is important is that the above
example shows the GW method works in the way as we anticipated.
Our major objective is to apply
GW-DMFT to strongly correlated electronic systems in higher
dimensions. we know E-DMFT works much better as is evident
from the scaling behavior with respect to the dimension. We also know
that the leading order perturbation in terms of the interaction
vertex, the GW, works better. We thus have a method which is much
easier to handle technically than C-DMFT and is able to
achieve the same goal to certain extent.\\

\section{Further Development and Outlook}
\label{sec-08}

E-DMFT allows to describe the interactions in a more flexible way
than we have presented so far. The need for such a freedom is evident
in the realistic calculation of materials, where, instead of the 
nearest neighbor repulsion considered in this paper we have to treat
the Coulomb interaction and its multipole expansion. For such a system,
the dielectric function is given by, in the linear response theory
\cite{mahan}:

\begin{equation}
\label{eq-63}
  \epsilon^{-1}(q,i\omega_n)=1+v_q \chi(q,i\omega_n)
\end{equation}

\noindent where $v_q$ is the Coulomb interaction given by, in 3D

\begin{equation}
\label{eq-64}
  v_q=\frac{4\pi e^2}{q^2}
\end{equation}

\noindent and $\chi(q,i\omega_n)$ is the density-density Green's
function defined through Eq.(\ref{eq-08}). One can make use of the
electron-phonon identity, Eq.(\ref{eq-10}) (set $\lambda=0$) and
get:

\begin{equation}
\label{eq-65}
  \epsilon(q,i\omega_n)=1-v_q \Pi(q,i\omega_n)
\end{equation}

\noindent Here the phonon self-energy $\Pi$ can be understood as
the collection of all the electron polarization diagrams. If one
proceeds with the formalism we presented in the previous sections,
the phonon self-energy $\Pi$ within E-DMFT approximation is
assumed to be momentum independent. However, Eq.(\ref{eq-65}) is
not the correct functional form for an insulator in which the
polarization should be
given by $\Pi(q,i\omega_n) \sim q^2f(q,\omega_n)$ with
$f$ weakly $q$ dependent.\\

To handle this situation in E-DMFT we have to tailor the formalism
to be compatible with the functional form of the response function.
This leads us to the following generalization of the action
discussed in Sections \ref{sec-02} and \ref{sec-03}:

\begin{equation}
\label{eq-66}
  S=S_0+\int_0^{\beta} d\tau \sum_{q} \left[ \frac{1}{2}
  \sum_{a,b} \phi_a (q,\tau) D^{-1}_{0,ab}(q) \phi_b (-q,\tau)
  -\sum_{a;\alpha,\beta} \phi_a (q,\tau)\rho_a(-q) \right]
\end{equation}

\noindent where we have defined a generalized electron density:

\begin{equation}
\label{eq-67}
  \rho_a(q)=\sum_k \sum_{\alpha\beta}c^{\dagger}_{k+q/2,\alpha}
  \Lambda_{a;\alpha\beta}(k+q/2,k-q/2) c_{k-q/2,\beta}
\end{equation}

\noindent with $\alpha$ and $\beta$ the spin labels. The indices $a$
and $b$ are used to label the local degrees of freedom other than
the spin, like the components of the multipole moments.
$S_0$ is the free action plus the local interaction. The key part here
is the electron-phonon vertex $\Lambda_{a;\alpha,\beta}(k+q/2,k-q/2)$
which is momentum dependent. A wise choice of this vertex allows to
preserve the physical momentum dependence in the response function
after the E-DMFT approximation.\\

We can define the following Green's function in the matrix form:

\[
  [D(q,\tau)]_{ab}=-\langle {\rm T}_{\tau}
  \phi_a (q,\tau) \phi_b (-q,0) \rangle
\]
\[
  [\tilde{\chi}(q,\tau)]_{ab}= -\langle {\rm T}_{\tau} 
  \rho_a(q,\tau)\rho_b(-q,0)\rangle
\]
\begin{equation}
\label{eq-68}
  =\sum_{k,k^{\prime}} \Lambda_{a,\alpha\beta}(k+q/2,k-q/2)
  \Lambda_{b,\alpha^{\prime}\beta^{\prime}}(k^{\prime}-q/2,k^{\prime}+q/2)
  \chi_{\alpha\beta,\alpha^{\prime}\beta^{\prime}}
  (k,q,\tau;k^{\prime},-q,0)
\end{equation}

\noindent with

\[
  \chi_{\alpha\beta,\alpha^{\prime}\beta^{\prime}}
   (k,q,\tau;k^{\prime},q^{\prime},0)=
  -\langle {\rm T}_{\tau}
  c^{\dagger}_{k+q/2,\alpha}(\tau) c_{k-q/2,\beta}(\tau)
  c^{\dagger}_{k^{\prime}+q^{\prime}/2,\alpha^{\prime}}(0)
  c_{k^{\prime}-q^{\prime}/2,\beta^{\prime}}(0)
   \rangle
\]

\noindent Same as before, we can derive an electron-phonon identity:

\begin{equation}
\label{eq-69}
  \left[\tilde{\chi}(q,i\omega_n)\right]^{-1}=-D_0(q)
  +\Pi^{-1}(q,i\omega_n),
\end{equation}

\noindent So far the results are exact. The E-DMFT approximation
amounts to mapping the general model (\ref{eq-66}) to an impurity
problem by integrating out all but one lattices site (we consider
the homogeneous phase here). The only new feature  is that
the general electron-phonon interaction vertex is not necessarily local
as we had before. The resulting phonon self-energy is a function of
frequency only. However, the general density-density Green's function
$\chi$, a physical quantity, now contains a non-trivial momentum
dependence which is evident from Eqs.(\ref{eq-68}) and (\ref{eq-69}).
Since the electron-phonon vertex $\Lambda$ can alway be
adjusted by redefining the auxiliary phonons, we are able to obtain
the desired momentum dependence from the physical considerations.\\

To see how it works, we go back to the example of the insulator.
We need the following form of the electron-phonon coupling in order
to describe the dipole-dipole interaction:

\begin{equation}
\label{eq-70}
  \Lambda_{a;\alpha\beta}(k+q/2,k-q/2)=
  \delta_{\alpha\beta}\frac{q_a}{q^2}
\end{equation}

\noindent with $a=x,y,z$. Meanwhile the free phonon propagator is of the
form

\begin{equation}
\label{eq-71}
  D_{0,ab}(q)=\frac{\delta_{ab}}{4\pi e^2},
\end{equation}

\noindent in order that the Coulomb interaction be recovered when the
auxiliary phonons are integrated out. Under this interaction vertex,
the auxiliary phonon represents an electric field mediating the
dipole-dipole interaction. Since only the longitudinal field is
coupled to the dipole moment, we can keep the corresponding phonon
mode and discard the transversal ones. For the E-DMFT
approximation, it is desired to work in the coordinate space where
the cavity construction is possible. To this end one needs to
convert the electron-phonon coupling to real space
which can be done by using the Wannier functions.\\

After solving the E-DMFT problem, the electron density Green's function
is given by

\begin{equation}
\label{eq-72}
  \chi(q,i\omega_n)=q^2\frac{-\Pi(i\omega_n)}{1+4\pi e^2 \Pi(i\omega_n)}
\end{equation}

\noindent and the dielectric function becomes

\begin{equation}
\label{eq-73}
  \epsilon(q,i\omega_n)=1+4\pi e^2 \Pi(i\omega_n),
\end{equation}

\noindent which has the correct form for an insulator.\\

To conclude, we have introduced in this section a way to tailor the
E-DMFT formalism so that the desired momentum dependence can
be preserved from physical considerations.\\

\section{Conclusion}
\label{sec-09}

In the paper we suggested a simple procedure of deriving E-DMFT
formalism, that is first separating out the Hartree contributions
and then making the E-DMFT approximation with regard to the fluctuations
around the Hartree ground state. This procedure is essential in the
phase with broken symmetry. It also helps to formulate the C-DMFT.\\

We developed an E-DMFT formulation by using a real
Hubbard-Stratonovich transformation. We introduced a local shift to
the general non-local interaction to ensure the positive-definiteness
of the effective interaction matrix. Our investigation showed that
in all the physical quantities the effects from the arbitrary shift
canceled out exactly. We also proved the equivalence of forming the
E-DMFT self-consistency by using the auxiliary phonon Green's
function and the two-electron Green's function. Based on these ideas,
we derived an E-DMFT of a single impurity site for a homogeneous
system with generic two particle interactions. We also presented a
formalism of E-DMFT combined with C-DMFT for a cluster of two lattice
sites, which is generalizable to clusters containing any number of
sites.\\

We suggested a generalized GW approach to incorporate the spatial
correlations into the E-DMFT approximation. While the on-site
self-energies are obtained non-perturbatively through E-DMFT,
those relatively weaker off-site contributions can be calculated
in a perturbative way. We identified the most important contributions 
to the non-local self-energies.\\

We showed how E-DMFT could be tailored to handle the response functions
with non-trivial momentum dependence in an insulator. Through the
example of the dielectric function we exhibited that an
appropriately defined electron-phonon vertex was able to keep the
correct functional form of the response function.\\

We implemented a QMC algorithm with shifts in E-DMFT to handle the
non-positive-definite interactions. This algorithm can be used for
a large variety of problems, including the Anderson lattice with
antiferromagnetic interactions. Two examples of the implementations
were presented.\\

The first example was the application of the single site E-DMFT
to the 3D U-V model. We studied the behavior of the electron
Green's function and the response function as the density instability
was approached. We studied the crossover between the metallic and the
Mott insulating phases. We investigated the frequency dependence
of the effective on-site interaction and showed its impact on the
single electron behavior. We showed a finite temperature phase
diagram of the 3D U-V model.\\

In the second example, we applied a single site E-DMFT combined with
GW method to a 1D U-V model with an alternating chemical potential.
It was found that the GW approach improved the E-DMFT results at
low frequencies in the desired direction. We also found that in the case
under investigation, it made little difference whether or not the
GW perturbation was performed within the E-DMFT iteration.\\

The success of the E-DMFT implementation opens the door to tackle
many complicated physical problems which could not be handled by simple
DMFT or other methods. The combination with GW and/or C-DMFT points
out a systematic way to improve the (E-) DMFT method.\\

\begin{acknowledgments}

This research was supported by NSF under Grant No. DMR-0096462 and by
the Center for Materials Theory at Rutgers University. We would like to
thank A. Georges, S. Florens, S. Savrasov, and V. Udovenko for useful
discussions and C.J. Bolech and S.S. Kancharla for the use of their
DMRG program.

\end{acknowledgments}

\begin{references}

\bibitem{kotliar1996}
A. Georges, G. Kotliar, W. Krauth, and M.J. Rozenberg,
Rev. Mod. Phys. {\bf 68}, 13 (1996).

\bibitem{dongen}
P.G.J. van Dongen,
Phys. Rev. Lett. {\bf 67}, 757 (1991);
Phys. Rev. B {\bf 49}, 7904 (1993);
{\it ibid}. {\bf 50}, 14016 (1994);
{\it ibid}. {\bf 54}, 1584 (1996).

\bibitem{schiller}
A. Schiller and K. Ingersent,
Phys. Rev. Lett. {\bf 75}, 113 (1995);
G. Zar{\'a}nd, D.L. Cox, and A. Schiller,
Phys. Rev. B {\bf 62}, R16227 (2000).

\bibitem{jarrell}
M.H. Hettler, A.N. Tahvildar-Zadeh, M. Jarrell, T. Pruschke,
H.R. Krishnamurthy,
Phys. Rev. B {\bf 58}, R7475 (1998);
M.H. Hettler, M. Mukherjee, M. Jarrell, and H. R. Krishnamurthy
Phys. Rev. B {\bf 61}, 12739 (2000).

\bibitem{georges}
S. Biermann, A. Georges, A. Lichtenstein, and T. Giamarchi,
Phys. Rev. Lett. {\bf 87}, 276405 (2001).

\bibitem{kotliar2001}
G. Kotliar, S. Savrasov, G. Palsson, and G. Biroli,
Phys. Rev. Lett. {\bf 87}, 186401 (2001) and the references therein.

\bibitem{kotliar2002}
G. Biroli and G. Kotliar,
Phys. Rev. B {\bf 65}, 155112 (2002).

\bibitem{subir}
S. Sachdev and J. Ye,
Phys. Rev. Lett. {\bf 70}, 3339 (1993).

\bibitem{kajueter}
H. Kajueter, Ph.D. thesis, Rutgers University (1996).

\bibitem{si2000}
Q. Si, S. Rabello, K. Ingersent, and J.L. Smith,
Nature {\bf 413} 804 (2001);
J.L. Smith and Q. Si,
Phys. Rev. B {\bf 61}, 5184 (2000);
Q. Si and J.L. Smith,
Phys. Rev. Lett. {\bf 77}, 3391 (1996).

\bibitem{georges1999}
O. Parcollet and A. Georges,
Phys. Rev. B {\bf 59}, 5341 (1999).

\bibitem{chitra2000}
R. Chitra and G. Kotliar,
Phys. Rev. Lett. {\bf 84}, 3678 (2000);
Phys. Rev. B {\bf 63}, 115110 (2001).

\bibitem{motome}
Y. Motome and G. Kotliar
Phys. Rev. B {\bf 62}, 12800 (2000).

\bibitem{pankov}
S. Pankov, G. Kotliar, and Y.Motome,
cond-mat/0112083.

\bibitem{anisimov}
V.I. Anisimov, A.I. Poteryaev, M.A. Korotin, A.O. Anokhin, and G. Kotliar,
J. Phys.: Condens. Matt. {\bf 9}, 7359 (1997).

\bibitem{savrasov2001}
S. Savrasov, G. Kotliar, and E. Abrahams,
Nature {\bf 410}, 793 (2001);
S. Savrasov and G. Kotliar,
Phys. Rev. Lett. {\bf 84}, 3670 (2000).

\bibitem{gabi}
G. Kotliar and S.Y. Savrasov,
{\it Model Hamiltonians and First Principles Electronic Structure
Calculations} in
{\it New Theoretical Approaches to Strongly Correlated Systems},
Ed. A.M. Tsvelik, Kluwer Academic Publishers (2001).

\bibitem{held}
For a review, see
K. Held, I.A. Nekrasov, G. Keller, V. Eyert, N. Bl{\"u}mer, A.K. McMahan,
R.T. Scalettar, T. Pruschke, V.I. Anisimov, and D. Vollhardt, in
{\it Quantum Simulations of Complex Many-Body Systems: From Theory
to Algorithms}, Ed. J. Grotendorst, D. Marx, and A. Muramatsu,
NIC Series Vol. 10 (NIC Directors, Forschungszentrum J{\"u}lich, 2002),
p. 175.

\bibitem{hirsch2001}
J.E. Hirsch,
Phys. Rev. Lett. {\bf 87}, 206402 (2001).

\bibitem{ferdi}
M. Springer and F. Aryasetiawan,
Phys. Rev. B {\bf 57}, 4364 (2001). 

\bibitem{zein}
N.E. Zein and V.P. Antropov,
cond-mat/0202483.

\bibitem{mahan}
G.D. Mahan,
{\it Many-Particle Physics,}
Plenum Press (1981),
Chapter 5.

\bibitem{hamann}
D.R. Hamann and S.B. Fahy,
Phys. Rev. B {\bf 47}, 1717 (1993).

\bibitem{hedin}
Lars Hedin,
Phys. Rev. {\bf 139} A796 (1965).

\bibitem{pietig}
R. Pietig, R. Bulla, and S. Blawid
Phys. Rev. Lett. {\bf 82}, 4046 (1999).

\bibitem{c-order1}
A. Ochiai, T. Suzuki, and T. Kasuya,
J. Phys. Soc. Jpn. {\bf 59}, 4129 (1990).

\bibitem{c-order2}
S. Mori, C.H. Chen, and S.-W. Cheong,
Nature (London) {\bf 392}, 473 (1998);
C.H. Chen and S.-W. Cheong,
Phys. Rev. Lett. {\bf 76}, 4042 (1996).

\bibitem{c-order3}
Tomioka, A. Asamitsu, H. Kuwahara, and Y. Tokurai,
J. Phys. Soc. Jpn. {\bf 66}, 302 (1997).

\bibitem{torrance}
J.B. Torrance, J.E. Vazquez, J.J. Mayerle, and V.Y. Lee,
Phys. Rev. Lett. {\bf 46}, 253 (1981);
J. B. Torrance, A. Girlando, J. J. Mayerle, J. I. Crowley,
V. Y. Lee, and P. Batail,
{\it ibid}. {\bf 47}, 1747 (1981).

\bibitem{nagaosa}
N. Nagaosa and J. Takimoto,
J. Phys. Soc. Jpn. {\bf 55}, 2735 (1986).

\bibitem{egami}
T. Egami, S. Ishihara, and M. Tachiki,
Science {\bf 261}, 1307 (1993);
S. Ishihara, T. Egami, and M. Tachiki,
Phys. Rev. B {\bf 49}, 8944 (1994);
S. Ishihara, M. Tachiki, and T. Egami,
{\it ibid}. {\bf 49}, 16123 (1994).

\bibitem{note1}
In E-DMFT, since the on-site interaction $U$ does not get renormalized,
one may find that the procedure of separating the Hartree contribution
does not affect the result as far as the local interaction is concerned.
So practically one can leave this term alone. The difference comes from
the non-local interaction where the bare interaction vertex is replaced
by an effective, on-site, and retarded one.

\bibitem{venky2}
C.J. Bolech, S.S. Kancharla, and G. Kotliar,
Preprint.

\bibitem{negele}
J.W. Negele and H. Orland,
{\it Quantum Many-Particle Systems,}
Addison-Wesley (1988), Chapter 2.

\bibitem{hirsch}
J.E. Hirsch and R.M. Fye,
Phys. Rev. Lett. {\bf 56}, 2521 (1986).

\bibitem{buendia1}
G.M. Buendia,
Phys. Rev. B {\bf 33}, 3519 (1986). 

\bibitem{note2}
The overall accuracy of the E-DMFT, which consists of the QMC
solution of the impurity model and the self-consist calculations,
is of the order $0.001$. If not otherwise specified, {\it all}
the data presented in this paper have the accuracy $\sim 0.001$.
The centers of the various symbols in the diagrams represent the
locations of the data while their sizes do {\it not} reflect the
accuracy. All the functions of the Matsubara frequency are
calculated at the corresponding discrete values. The lines
connecting the points are guides to the eye.

\bibitem{note3}
In presenting the phonon related results we
{\it always} set $\lambda=0$. A finite $\lambda$ is used {\it only}
in the QMC simulations.

\bibitem{doniach}
S. Doniach and E.H. Sondheimer,
{\it Green's Functions for Solid State Physicists,}
Benjamin (1974), Chapters 6 and 7.

\bibitem{hanke}
G. Strinati, H.J. Mattausch, and W. Hanke,
Phys. Rev. Lett. {\bf 45}, 290 (1980);
Phys. Rev. {\bf B} 25, 2867 (1982).

\bibitem{holm}
B. Holm and U. von Barth,
Phys. Rev. B {\bf 57} 2108 (1998).

\bibitem{leeuwen}
C.-O. Almbladh, U. Von Barth, and R. Van Leeuwen,
Int. J. Mod. Phys. B {\bf 13}, 535 (1999).

\bibitem{abrikosov}
A.A. Abrikosov, L.P. Gorkov, I.E. Dzyaloshinski,
{\it Methods of Quantum Field Theory in Statistical Physics,}
Prentice-Hall (1963), Section 10.

\bibitem{nozieres}
P. Nozieres,
{\it Theory of Interacting Fermi Systems,}
Addison-Wesley (1964), Chapter 3.

\bibitem{metzner}
W. Metzner and D. Vollhardt,
Phys. Rev. Lett. {\bf 62}, 324 (1989).

\bibitem{godby}
L. Steinbeck, A. Rubio, L. Reining, M. Torrent, I.D. White, and R.W. Godby,
Comput. Phys. Commun. {\bf 125}, 105 (2000);
T.J. Pollehn, A. Schindlmayr, and R.W. Godby,
J. Phys.: Cond. Matt. {\bf 10}, 1273 (1998).

\bibitem{venky0}
S.S. Kancharla and C.J. Bolech,
Phys. Rev. B {\bf 64}, 85119 (2001).

\bibitem{venky1}
We would like to thank Dr. S. S. Kancharla and Dr. C. J. Bolech
for allowing us to use their DMRG programs for the calculation.

\bibitem{torio}
M.E. Torio, A.A. Aligia, and H.A. Ceccatto
Phys. Rev. B {\bf 64}, R121105 (2001) and the references therein.

\end{references}
\end{document}